\documentclass[twocolumn]{aastex63}

\usepackage{float}
\usepackage{amsmath}
\usepackage{comment}
\usepackage{graphicx}
\usepackage[caption=false]{subfig}

\submitjournal{ApJ}

\graphicspath{{./}{figures/}}

\begin{document}

\title{Spatially Resolved Modeling of Optical Albedos for a Sample of Six Hot Jupiters}

\shorttitle{Spatially Resolved Modeling of Optical Albedos}
\shortauthors{Adams et al.}

\correspondingauthor{Danica Adams}
\email{djadams@caltech.edu}

\author[0000-0001-9897-9680]{Danica J. Adams}
\affiliation{Division of Geological and Planetary Sciences, California Institute of Technology, Pasadena, CA 91125, USA}

\author[0000-0003-3759-9080]{Tiffany Kataria}
\affiliation{Jet Propulsion Laboratory, California Institute of Technology, Pasadena, CA 91109, USA}

\author[0000-0003-1240-6844]{Natasha E. Batalha}
\affiliation{NASA Ames Research Center, MS 245-3, Moffett Field, CA 94035, USA}

\author[0000-0002-8518-9601]{Peter Gao} \affiliation{Department of Astronomy and Astrophysics, University of California, Santa Cruz, Santa Cruz, CA 95064, USA}\affiliation{Earth and Planets Laboratory, Carnegie Institution for Science, 5241 Broad Branch Road, NW, Washington, DC 20015, USA}

\author[0000-0002-5375-4725]{Heather A. Knutson}
\affiliation{Division of Geological and Planetary Sciences,
California Institute of Technology, Pasadena, CA 91125, USA}

\begin{abstract}

Optical secondary eclipse measurements made by \emph{Kepler} reveal a diverse set of geometric albedos for hot Jupiters with equilibrium temperatures between $1550-1700$ K. The presence or absence of high altitude condensates, such as Mg$_2$SiO$_4$, Fe, Al$_2$O$_3$, and TiO$_2$, can significantly alter optical albedos, but these clouds are expected to be confined to localized regions in the atmospheres of these tidally locked planets. Here, we present 3D general circulation models and corresponding cloud and albedo maps for six hot Jupiters with measured optical albedos in this temperature range. We find that the observed optical albedos of K2-31b and K2-107b are best matched by either cloud free models or models with relatively compact cloud layers, while Kepler-8b and Kepler-17b's optical albedos can be matched by moderately extended ($f_{sed}$ = 0.1) parametric cloud models. HATS-11b has a high optical albedo, corresponding to models with bright Mg$_2$SiO$_4$ clouds extending to very low pressures ($f_{sed}$ = 0.03).  We are unable to reproduce Kepler-7b's high albedo, as our models predict that the dayside will be dominated by dark Al$_2$O$_3$ clouds at most longitudes. We compare our parametric cloud model with a two-zone microphysical cloud model (\texttt{CARMA}). We find that even after accounting for the 3D thermal structure, no single cloud model can explain the full range of observed albedos within the sample. We conclude that a better knowledge of the vertical mixing profiles, cloud radiative feedback, cloud condensate properties, and atmospheric metallicities is needed in order to explain the unexpected diversity of albedos in this temperature range. 

\end{abstract}



\section{Introduction} \label{sec:intro}

Transiting short period gas giant planets, or `hot Jupiters', are among the most favorable targets for atmospheric characterization studies. By observing the decrease in optical flux when the planet passes behind the host star (secondary eclipse), we can measure the optical dayside albedos for these tidally locked planets. Most of these albedo measurements have come from space telescopes observing in broad optical bandpasses \citep[e.g.][]{rowe2008,kipping2011,demory2011,demory2013,coughlin2012,parviainen2013,angerhausen2014,vonparis2016,niraula2018}. The hot Jupiter HD~189733b is currently the only planet with a spectroscopically resolved reflected light measurement \citep[][geometric albedo of $0.40\pm0.12$ at $290-450$ nm and $<0.12$ at $450-570$ nm]{evans2013}.  WASP-12b and WASP-43b were also observed with optical spectrographs on \emph{HST}, but both observations resulted in non-detections \citep[][geometric albedos $<0.06$]{bell2017,fraine2021}.  These spectroscopic albedo measurements are invaluable for constraining the nature of the scattering particles in the atmospheres of these planets \citep[e.g.][]{barstow2014}.   

\begin{figure}
    \includegraphics[width=0.5\textwidth]{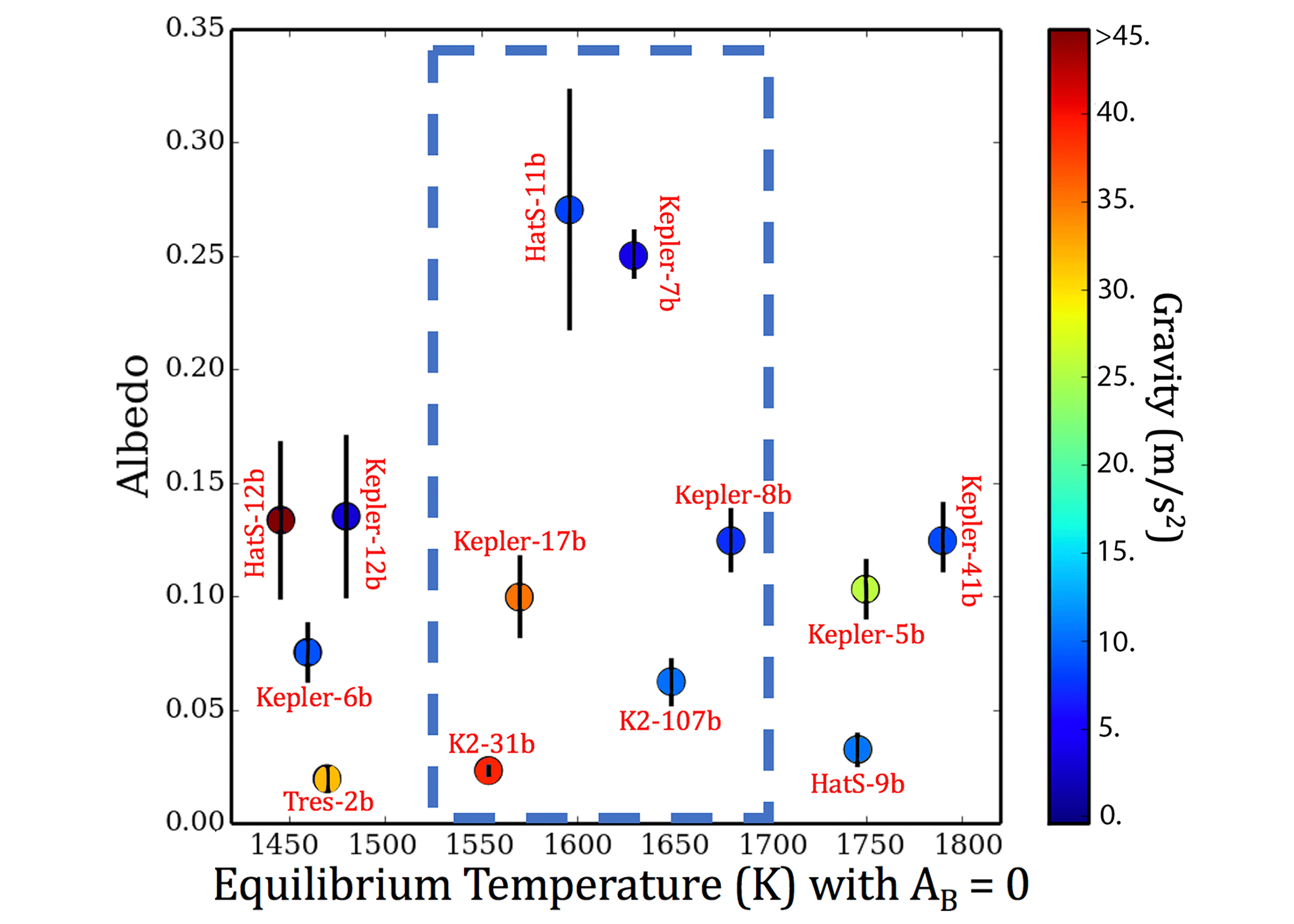}
    \caption{Optical geometric albedo measurements for a sample of hot Jupiters observed in the \emph{Kepler} bandpass; at these temperatures thermal emission is negligible for planets with cloudy skies and the measured secondary eclipse depth is dominated by reflected light. The equilibrium temperature is calculated assuming a Bond albedo of zero and efficient day-night circulation. The planet colors vary as a function of surface gravity. The dashed box indicates the temperature range considered in this study. Measurements are drawn from \citet[][]{fortney2011, desert2011, barclay2012, esteves2013, shporer2015, niraula2018, heng2020}.
    }
    \label{fig:kepalbedos}
\end{figure}

Theoretical models predict that variations in hot Jupiter optical albedos should primarily be driven by the presence or absence of high altitude aerosols, which are expected to scatter incident starlight \citep[e.g.][]{seager2000,burrows2008}. Cloud-free atmosphere models for hot Jupiters generally predict dayside albedos of less than 0.1 \citep{seager2000}, but the observed optical geometric albedo measurements published to date span a wide range of values, with the brightest planets exhibiting albedos as high as $0.3$ \citep[e.g.][]{Heng_2013,niraula2018}.  In the solar system, the presence of water clouds increases Earth's geometric albedo to approximately 0.37 \citep[e.g.][]{goode2001}, while ammonia clouds in Jupiter's atmosphere contribute to its geometric albedo of \textcolor{black}{approximately 0.5} \citep[e.g.][]{marley1999}.  For the same reason, models suggest that the presence of reflective condensates such as Mg$_2$SiO$_4$ or MgSiO$_3$ in hot Jupiter atmospheres can increase their albedos to values as high as 0.5 \citep{parmentier2016,parmentier2021,roman2021}.  Unlike brown dwarfs, whose cloud distributions and optical depths correlate closely with their equilibrium temperatures \citep[e.g.][]{kirkpatrick2005,marley2010}, hot Jupiter albedos can vary by as much as an order of magnitude within a relatively narrow range of equilibrium temperatures (Figure \ref{fig:kepalbedos}). This suggests that individual planets with similar equilibrium temperatures may exhibit diverse cloud properties.

Hot Jupiters are expected to be tidally locked as a result of their short orbital periods, which can lead to significant day-night temperature gradients \citep[e.g.][]{guillot2002}.  This inhomogneous temperature structure affects the cloud distribution, and 3D atmospheric circulation models predict that hot Jupiters should host spatially inhomogenous clouds over a wide range of equilibrium temperatures \citep[e.g.][]{parmentier2016, parmentier2021, roman2021}. Furthermore, there is direct observational evidence for inhomogeneous cloud structures stemming from reflected-light phase curves \citep[e.g.,][]{demory2013}.  To date, three of the four planets with observed reflected-light phase curves appear to have patchy clouds \citep[][also see Figure \ref{fig:kepalbedos}]{desert2011,demory2013, angerhausen2014,esteves2015,shporer2015,vonparis2016,niraula2018}. Kepler-7b, -12b, and -41b all appear to have patchy clouds that are preferentially concentrated in the western (dawn) hemisphere, causing the peak of the phase curve to occur after the secondary eclipse \citep{shporer2015}. Observations of the fourth planet, TrES-2b, indicate that it is uniformly dark ($A_s<0.03$, which is equivalent to $A_g<0.02$ for a Lambertian sphere, where $A_s$ is the spherical albedo and $A_g$ is the geometric albedo) at all phases \citep{vonparis2016}. These studies demonstrate that if we wish to explain the observed planet-to-planet variations in the measured dayside optical albedos of hot Jupiters, we must utilize 3D models capable of capturing the spatially varying cloud structure. 

While an increasing number of studies are using general circulation models (GCMs) to predict cloud patterns, these have either focused on individual planets \citep[e.g.][]{webber2015,lee2015,oreshenko2016,helling2019a,helling2019b,helling2020,lines2018,lines2019,roman2019} or generic grids of models \citep[e.g.][]{parmentier2016,parmentier2021,roman2021}. Of these listed studies, only five \citep{webber2015,oreshenko2016,parmentier2016,parmentier2021,roman2021} use their models to calculate predicted albedos in the \emph{Kepler} bandpass. The question of why individual planets with similar equilibrium temperatures would exhibit widely varying cloud properties therefore remains \textcolor{black}{largely} unexplored in the literature.

\begin{centering}
\begin{deluxetable*}{lccccccccc}[!t]
\tabletypesize{\small}
\tablecaption{Properties of our planet sample.}
\tablewidth{0pt}
\tablehead{
\colhead{Planet}&\colhead{$M_P$ ($M_J$)} &\colhead{$R_P$ ($R_J$)}&\colhead{${T_{eq}}  (K)$\tablenotemark{a}}&\colhead{$g$ $(m/s^2)$}&\colhead{a (AU)}& \colhead{Period (days)} & \colhead{Measured $A_g$} & \colhead{Ref.$^1$}}
\startdata
K2-31b & 1.77 & 1.06 & 1550 & 39.07 & 0.022 & 1.26 & 0.023$\pm$0.002 & 1,\textbf{9} \\
Kepler-17b & 2.45 & 1.31 & 1570 & 35.41 &  0.026 & 1.49 & 0.099$\pm$0.017 & 2,\textbf{3} \\
HATS-11b & 0.85 & 1.51 & 1560 & 7.97 & 0.051 & 3.62 & 0.270$\pm$0.052 & 4,5,\textbf{9} \\
Kepler-7b & 0.45 & 1.65 & 1630 & 4.16 & 0.062 & 4.89 & \textcolor{black}{0.25$\pm$0.01} & 6,\textbf{8} \\ 
K2-107b & 0.84 & 1.44 & 1650 & 10.26 & 0.048 & 3.31 & 0.062$\pm$0.010 & 5,7,\textbf{9} \\
Kepler-8b & 0.60 & 1.41 & 1680 & 7.32 & 0.047 & 3.85 & 0.124$\pm$0.013 & 6,\textbf{10}
\enddata 
\label{table:systems1}
\tablenotetext{1}{Reference to geometric albedo measurement is bolded for each row.}

\tablenotetext{a}{Calculated assuming a Bond albedo of zero and efficient day-night recirculation.}
\tablecomments{(1) \citet{grziwa2016}, (2) \citet{desert2011}, (3) \citet{bonomo2017}, (4) \citet{bayliss2018}, (5) \citet{livingston_2018}, (6) \citet{esteves2015}, (7) \citet{eigmuller2017}, (8) \citet{heng2020}, (9) \citet{niraula2018}, (10) \citet{esteves2013}}
\end{deluxetable*}
\end{centering}

Modeling the 3D structure of clouds can be computationally demanding, especially given the large number of parameters that must be considered (including but not limited to the particle number density and size distribution, spatial extent of the clouds, location of the cloud decks, number and composition of cloud species, and coupled radiative feedback). In examining the body of published studies that use GCMs to predict hot Jupiter dayside albedos, we find that all of these studies make simplifying assumptions in order to conserve run time and numerical complexity. \citet{oreshenko2016}, for example, determined the locations of clouds by comparing the 3D thermal structure from a GCM with relevant condensation curves, but assumed a fixed particle size and local condensation only (i.e., they neglected vertical mixing) for the cloud layers. \textcolor{black}{In \citet{parmentier2016}, cloud distributions were determined using the completed temperature structure calculated by a non-gray cloud-free GCM, with a cloud-top minimum pressure fixed at 1 microbar.  In \citet{parmentier2021}, cloud layers were also calculated using the thermal structure output from a GCM, but the prescribed vertical extent is limited by temperatures within the extent of 200 mbar to 1 $\mu$bar. \cite{roman2019} and \cite{roman2021} differ significantly from the Parmentier models. In the former, cloud distributions were determined at each timestep in a double-grey GCM which included radiative feedback. They included extended clouds which form when the temperature-pressure profile permits, but are forced to taper off at pressures between 0.3 mbar and 0.057 mbar. They also consider compact cases with varied optical thickness where the clouds are truncated and tapered off after approximately one scale height regardless of where their base forms. In \cite{roman2021}, they also consider a more extensive grid of planet models, while also varying cloud compositions, densities, and vertical extents. All of the studies listed above assumed homogeneous condensation, allowing them to treat each individual cloud species separately. Although there are studies in the literature that have combined GCMs with microphysical cloud models \citep[e.g.,][]{lee2015,lines2018,helling2019a,helling2019b,helling2020}, these models focused on individual planets and were limited in their albedo predictions in the \emph{Kepler} bandpass.} \color{black}{Although there are studies in the literature that have combined GCMs with microphysical cloud models \citep[e.g.,][]{lee2015,lines2018,helling2019a,helling2019b,helling2020}, these models focused on individual planets and were limited in their albedo predictions in the \emph{Kepler} bandpass.}

In this study, we utilize a suite of models to investigate the role of patchy clouds over a sample of six individual hot Jupiters, chosen due to their diverse observed albedos over a narrow range of equilibrium temperatures. In \S\ref{sec:2methods}, we describe our sample selection and summarize our modeling approach.  We use 3D GCMs to derive the thermal structure and eddy diffusion coefficients ($K_{zz}$). We then use \texttt{Virga}, a phase equilibrium cloud code, to make detailed maps of cloud structure over the dayside of each planet. We use \texttt{PICASO}, a radiative transfer program, to calculate the corresponding geometric albedo maps and hemisphere-integrated dayside albedos. We also consider a microphysical code, \texttt{CARMA}, which enables us to take a closer look at the role of nucleation, condensation, and sedimentation in shaping the distribution of dayside clouds. 
In \S\ref{sec:results} we compare our predicted dayside-integrated optical albedos with observations of the six planets of interest. We then investigate the relative importance of various model assumptions, such as equilibrium condensation versus kinetic condensation, by comparing the \texttt{Virga} results with the \texttt{CARMA} results. Finally, we discuss the implications of our results in \S\ref{sec:discussion} and present our conclusions in \S\ref{sec:conclusions}.

\section{Methods}\label{sec:2methods}

\begin{figure*}[!htb]
    \centering
    \includegraphics[width=0.8\textwidth]{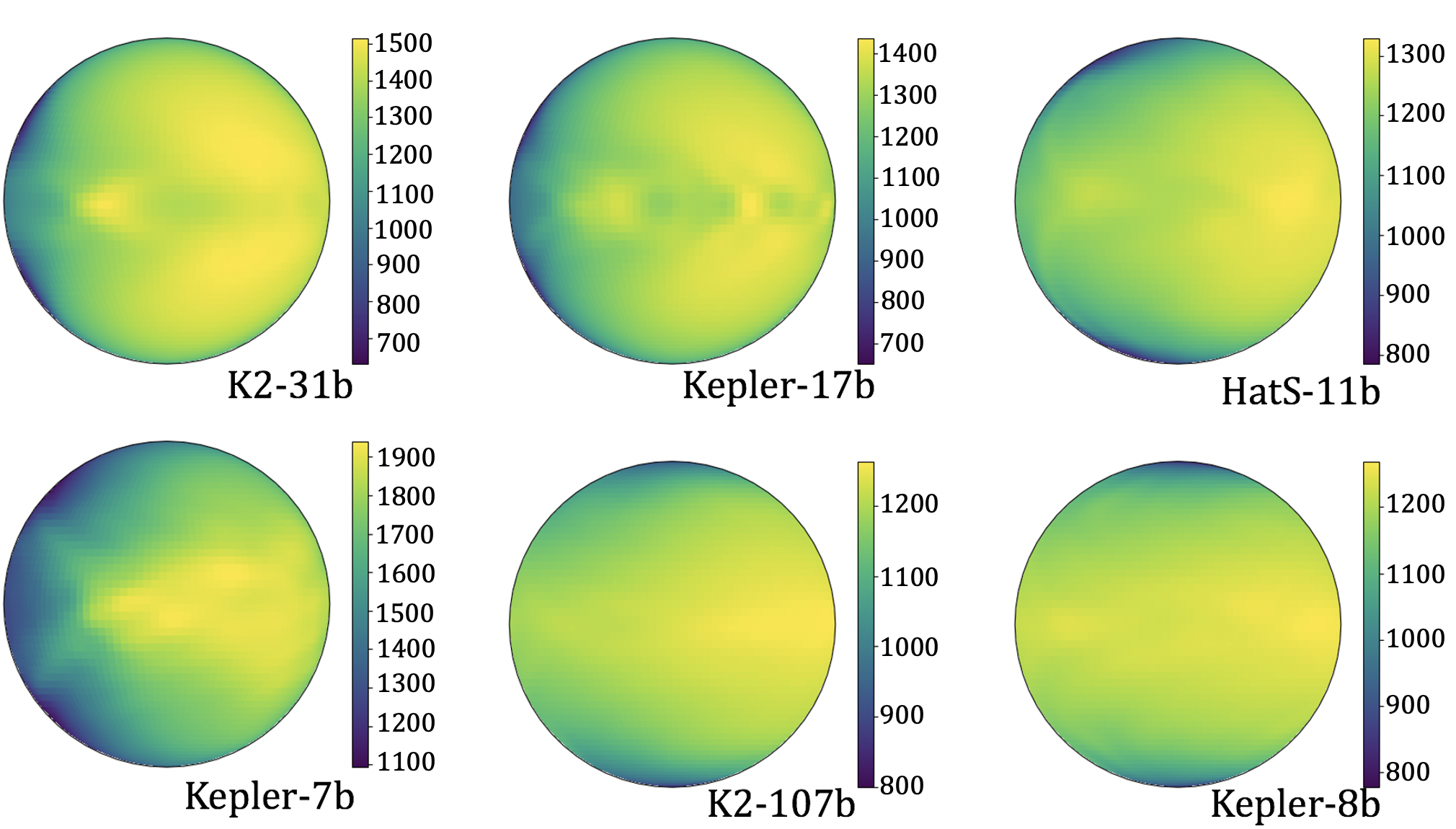}
    \caption{Map of the dayside temperatures (in Kelvin) of each planet at 1 mbar, roughly the pressure of unit optical depth in a clear atmosphere in the \emph{Kepler} bandpass. Each planet is given a unique scale for the color bar to best match the relevant temperature range.}
    \label{fig:tmaps}
\end{figure*}

\subsection{Planet Sample}\label{sec:sample}

In this study we focus on planets with equilibrium temperatures ($T_{eq}$, calculated assuming an albedo of zero and efficient day-night recirculation) between $1550-1700$ K.  For planets in this relatively narrow temperature range, we expect that reflective silicate clouds should dominate the optical dayside albedos \citep[e.g.][]{parmentier2016,powell2018,parmentier2021,gao2020,roman2021}. This temperature range contains some of the most reflective hot Jupiters observed to date, including Kepler-7b \citep{demory2013,heng2020} and HATS-11b \citep{niraula2018}. Note that two observed geometric albedos have been reported for Kepler-7b, and from here on we consider the most recent value from \citet{heng2020}.  It also includes three moderately reflective hot Jupiters \citep[Kepler-8b, Kepler-17b, and K2-107b;][]{esteves2013,desert2011,niraula2018} and one very dark hot Jupiter \citep[K2-31b;][]{niraula2018}.  Previous studies \citep[e.g.,][]{demory2013} have concluded that planets with these equilibrium temperatures should have negligible amounts of thermal emission in the \emph{Kepler} bandpass.  We checked this by using \texttt{PICASO} to compute the predicted thermal emission for each of the six planets in our sample and found a contribution of 2 ppm or less.  This is significantly smaller than the uncertainties on the secondary eclipse depths used to calculate the geometric albedos for these planets. Hence, throughout this paper we ignore the thermal contribution to the measured secondary eclipse depth in the \emph{Kepler} bandpass.

In this study we present a GCM tailored to each of the six individual planets, which we use to predict planet-to-planet variations in cloud coverage and dayside albedo in the optical \emph{Kepler} band. These planets sample a range of surface gravities, allowing us to investigate its effect on the planet's thermal structure, cloud distributions, and optical albedo. We summarize the physical and orbital properties of each system in Table \ref{table:systems1}.


\subsection{Modeling Atmospheric Circulation with the SPARC/MITgcm}\label{sec:gcm}

We model each planet’s \textcolor{black}{clear-sky (cloud-free)} 3D thermal structure and atmospheric circulation with the Substellar and Planetary Radiation and Circulation (SPARC) model, which couples the GCM maintained at the Massachusetts Institute of Technology \citep[the MITgcm;][]{adcroft2004} with a plane-parallel, two-stream version of the multi-stream radiation code as described in \citet{marley1999}. The MITgcm solves the 3D primitive equations on a staggered Arakawa C grid \citep{arakawa1977} with the finite-volume method. The equations are discretized on a 128 $\times$ 64 cubed-sphere grid with 53 vertical layers extending from 200 bars at the bottom boundary to 20 $\mu$bar at the top boundary. A horizontal fourth-order Shapiro filter is used to smooth horizontal noise. \color{black}{We let each model run for a simulated 1000+ Earth days so as to reach quasi-steady state equilibrium}. 

\color{black}The radiative transfer scheme solves the two-stream radiative transfer equations using the correlated-k method \citep[][]{goody1989,marley1999} over 11 spectral bins \citep[][]{kataria2013}. This coupling allows for the self-consistent calculation of the heating and cooling rates of the atmosphere with latitude, longitude and pressure. At each grid point, the radiative transfer scheme calculates the upward and downward fluxes at each pressure layer, which are used to update the heating/cooling rates. These rates are used by the MITgcm to update the wind and temperature fields. Opacities are computed at each pressure-temperature point assuming chemical and thermodynamic equilibrium, using the solar photospheric elemental abundances of \citet{lodders03}. We interpolate across the PHOENIX stellar atmosphere models to generate an input spectrum for each host star. The SPARC/MITgcm has been successfully utilized for a series of hot Jupiter studies \citep[e.g.][]{showman2009,showman2013,showman2015,kataria2013,kataria2015,kataria2016,parmentier2013,lewis2014}, and we refer the reader to \citet{kataria2016} for further details. Figure \ref{fig:tmaps} shows the resulting thermal structure for each of the six planets in our sample at 1 mbar (approximately the level of unit optical depth for clear skies).  

\subsection{Computing Equilibrium Condensate Clouds with \texttt{Virga}}\label{sec:virga}

We use the thermal structure and vertical mixing rates from the SPARC/MITgcm model as inputs to \texttt{Virga} \citep{virga}\footnote{Code and documentation available at https://natashabatalha.github.io/virga/}, an open-source code that calculates phase equilibrium cloud distributions.  The cloud parameterization used in this code is described in \citet{ackerman2001} and has been used for 1D model studies across a wide range of exoplanets and brown dwarfs \citep[e.g.][]{fortney2006,marley2010,morley2015}. This parametric approach also allows us to sample the 3D cloud structure at a much higher spatial resolution than for our microphysical models. 

We calculate $K_{zz}$ profiles from the root-mean-square (rms) vertical velocities derived from global horizontal averages at a given pressure level from the GCMs by assuming $K_{zz} = w(z)L(z)$, where $w(z)$ is the horizontally averaged global rms vertical velocity from the GCM simulations,  $L(z)$ is approximated as the atmospheric pressure scale height $H(z)$ \citep[but could be a fraction of $H(z)$; see][]{smith1998}, and $z$ is altitude. \textcolor{black}{Moses et al. (2011) note that this is only an estimate; a better approach would involve calculating $K_{zz}$ from the eddy vertical velocity times the eddy displacement, but this information is not readily obtainable from the GCMs; this could be resolved by adding passive tracers to future GCM models.} Our treatment may overestimate $K_{zz}$ in the $\sim$10–200 bar radiative region, where the vertical motion often consists of small-scale wave oscillations. 

\textcolor{black}{We reduce the spatial resolution of our longitude and latitude grid from 128$\times$65 to 10$\times$10 by binning the pressure-temperature profiles and corresponding Kzz profiles prior to running Virga. We retain the original 53 pressure levels in the rebinned grid.} 
This binning has a negligible effect on our calculation of the phase-integrated albedo and significantly reduces computation time. We bin using the area mean with angles from the Chebyshev-Gauss integration method that vary as a function of planetary latitude and longitude. 

In \texttt{Virga} the molar mixing ratio of the condensed phase, $q_c$, is calculated by solving the equation:

\begin{equation}
    -K_{zz} \frac{\partial q_t}{\partial z} - f_{sed} \omega_* q_c = 0
    \label{eq:eddysed}
\end{equation}
where $q_t$ is the total mixing ratio (condensed and vapor phases), $\omega_*$ is the convective velocity scale, and $f_{sed}$ is defined as the ratio of the mass-weighted droplet sedimentation velocity to  the convective velocity, $\omega_*$. The product $f_{sed} \omega_*$ describes an average sedimentation velocity for the condensate, which offsets turbulent mixing.  We refer the reader to \citet{ackerman2001} for more details regarding the equations that govern \texttt{Virga}.

In Eq. \ref{eq:eddysed}, $f_{sed}$ is the only parameter that cannot be calculated directly from the models.  We therefore treat it as a tunable parameter and explore a range of possible values. Models with larger values of $f_{sed}$ will have high rates of sedimentation, concentrating the condensing species in the lower atmosphere. Conversely, models with smaller values of $f_{sed}$ will have much slower sedimentation rates, allowing cloud particles to remain lofted higher in the atmosphere. 
For each planet, we run a suite of \texttt{Virga} models with $f_{sed}$ values of 0.03, 0.1, 0.3, 1.0, 3.0, and 6.0. This range is motivated by comparisons to observational data, which suggest that $f_{sed}$ can be as small as 0.01 for super Earths \citep[][]{morley2015} or as large as 2-5 for some gas giants and brown dwarfs \citep{skemer2016,macdonald2018,saumon2008}. For Jupiter's ammonia clouds, an $f_{sed}$ of $\sim$2 appears to provide the best match to observations \citep[][]{ackerman2001}. We therefore conclude that our chosen list of values spans a representative range for this parameter.

\subsection{Computing Microphysical Clouds with \texttt{CARMA}}\label{sec:carma}

\begin{figure}[t]
    \centering
    \includegraphics[width=3.2in]{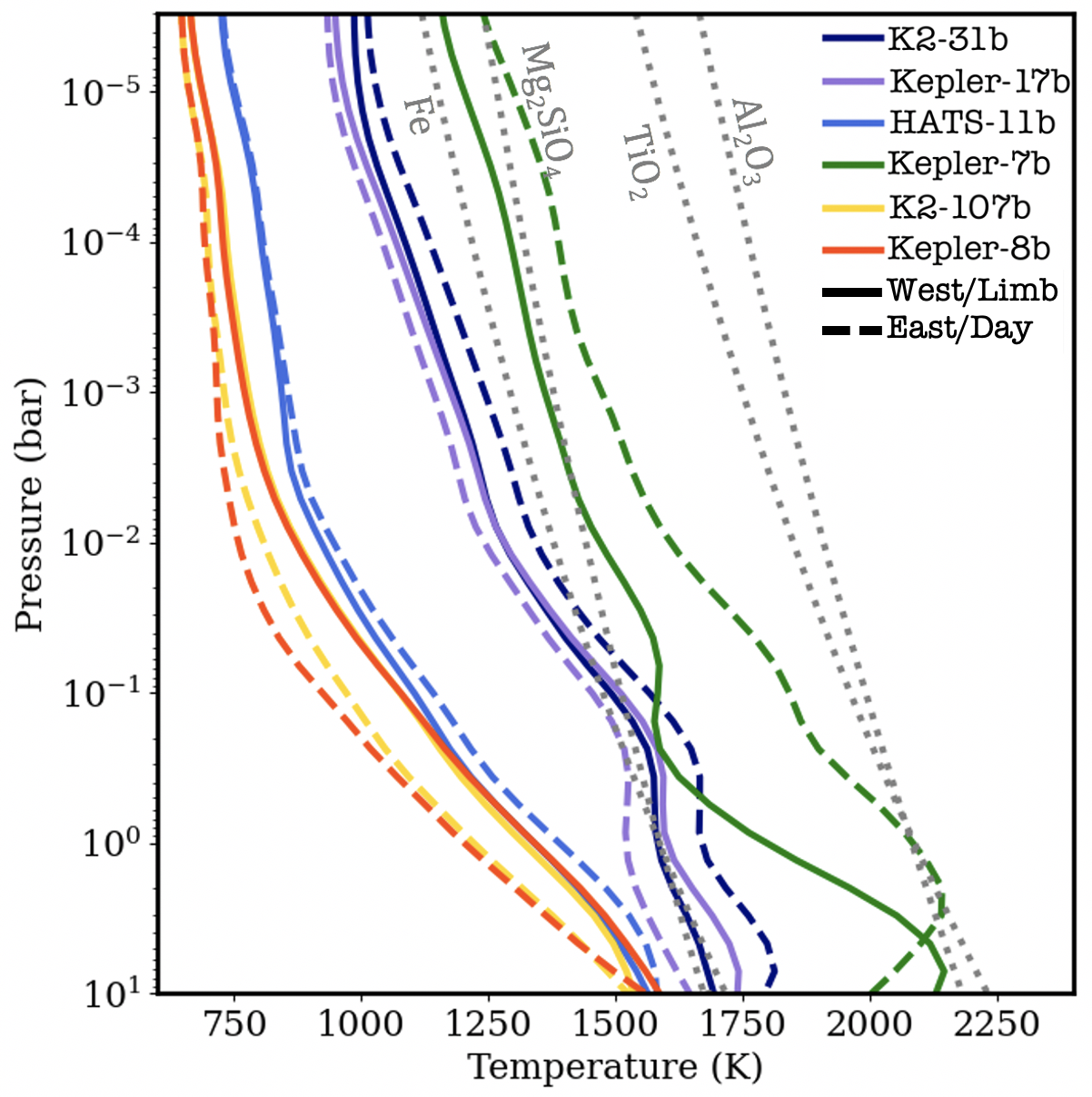}
    \caption{Condensation curves (dashed grey) of Fe, Al$_2$O$_3$, TiO$_2$, and Mg$_2$SiO$_4$ clouds compared with the two-zone model averaged temperature profiles (see \S\ref{sec:carma_comparison}). Dayside (solid) and western limb (dashed) are shown for each planet: K2-107b (yellow), Kepler-8b (orange), HATS-11b (black), K2-31b (navy), Kepler-17b (purple), and Kepler-7b (green).}
    \label{fig:TCondens}
\end{figure}

In addition to the parametric model described above, we also utilize the Community Aerosol and Radiation Model for Atmospheres (\texttt{CARMA}), a more computationally demanding microphysical cloud formation model.  \texttt{CARMA} calculates the equilibrium cloud particle size distribution by solving the 1D discretized continuity equation for aerosol particles that experience vertical transport due to sedimentation and eddy diffusion and production and loss due to particle nucleation (homogeneous and heterogeneous), condensation, evaporation, and coagulation. \texttt{CARMA} has been previously used to investigate condensate cloud formation on Earth \citep[e.g.,][]{ackerman1993,ackerman1995,jensen1994}, Venus \citep[e.g.][]{james1997,mcgouldrick2007,GAO201483}, Mars \citep[e.g.][]{colaprete1999}, and exoplanets \citep[e.g.,][]{gao2018,gao2020,Powell_2019}.  \texttt{CARMA} has also been used to model photochemical hazes on Titan \citep{toon1992}, Pluto \citep{gao2017pluto}, ancient Earth \citep{wolf2010}, and warm Jupiters \citep{adams2019}. In order to make our use of \texttt{CARMA} computationally tractable, we divide each planet into two zones and calculate averaged temperature and $K_{zz}$ profiles for each zone as described in \S\ref{sec:carma_comparison} (see also Figure \ref{fig:TCondens}).  We do not consider photochemical hazes here, as the planets in our sample lie above the temperature range where these hazes are expected to form \citep{lodders2002,kawashima2019,gao2020}.


In the \texttt{CARMA} model, the formation of condensate clouds begins with either homogeneous or heterogeneous nucleation. Cloud particles undergo homogeneous nucleation when stable clusters of condensate molecules form and grow directly from the vapor. The rate is controlled by the material properties of the condensate, such as its molecular weight and surface energy, and the flux of molecules to the cluster, which depends on the abundance of condensate vapor. Under the same supersaturation and local temperature, high surface energy and molecular weight materials tend to nucleate more slowly than low surface energy and low molecular weight materials. Unlike homogeneous nucleation, heterogeneous nucleation involves the formation of stable clusters on condensation nuclei, or foreign surfaces, which are provided by other aerosol particles in the atmosphere. The size and abundance of these particles strongly impact the rate of heterogeneous nucleation. The nucleation rate is also dependent on the interaction between the condensate and the surface, characterized by the contact angle between the surface and the condensate cluster, the energy needed by a condensate molecule to desorb from the surface, and the oscillation frequency of the condensate molecule on the surface, which is related to the desorption energy \citep[][]{Pruppacher1978}. 

 
Unlike in \texttt{Virga}, which assumes a log-normal particle size distribution, particle size distributions in \texttt{CARMA} are resolved using mass bins and can change over time via condensation, evaporation, and coagulation. The former two rates are dependent on the flux of condensate molecules and the rate at which particles may conduct away the latent heat released upon condensation. Coagulation, or growth via physically sticking upon the collision of particles, is also considered. Each mass bin corresponds to particle masses twice that of the previous bin. We use 65 bins in our model, with the mass in the first bin corresponding to particles with radii of 0.1 nm for all species.


\subsection{Cloud Compositions}\label{sec:cloudcomp}

\begin{figure*}
    \centering
    \includegraphics[width=7in]{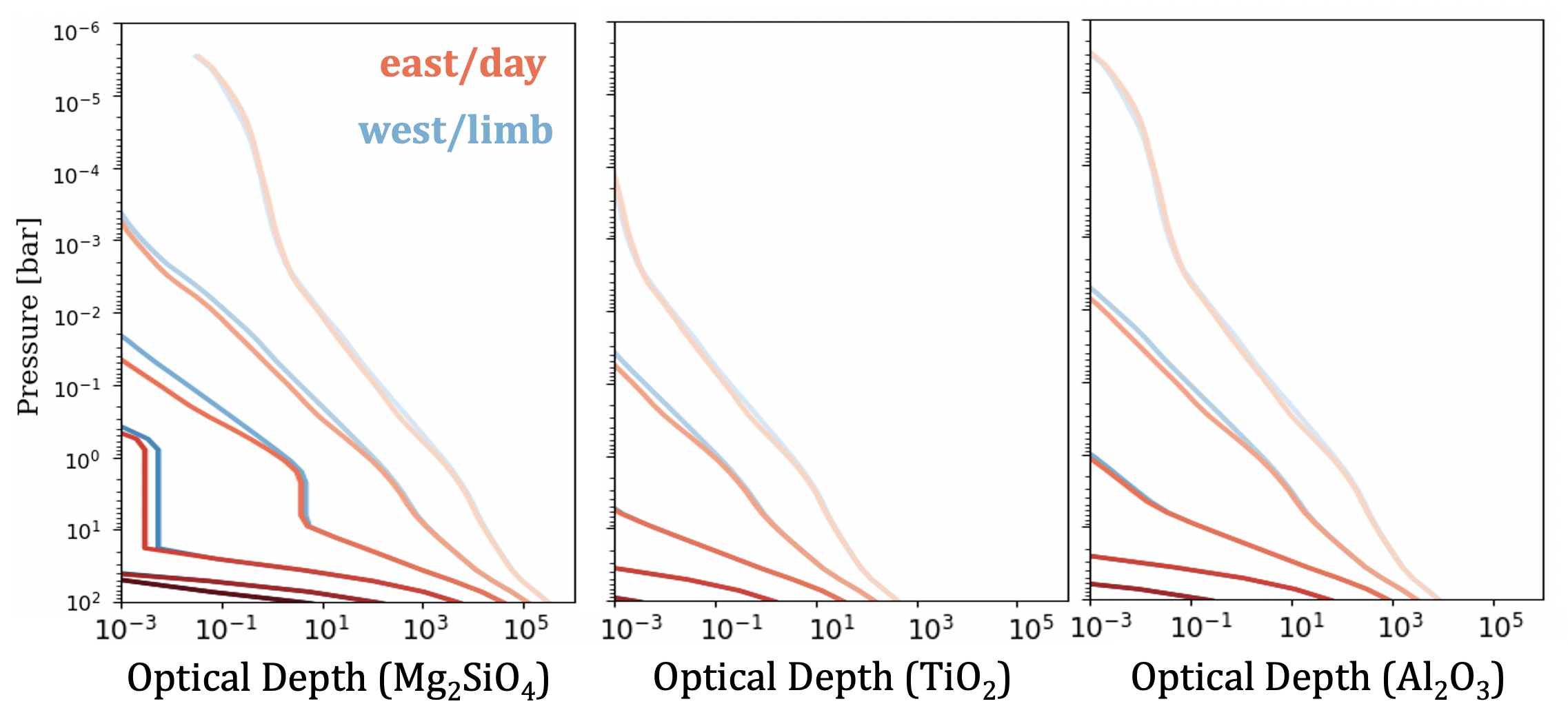}
    \caption{Nadir optical depths (integrated over the Kepler bandpass) for three condensate species in our \texttt{Virga} model of K2-107b. From left to right: Mg$_2$SiO$_4$, TiO$_2$, Al$_2$O$_3$. Each panel shows two representative grid points: black corresponds to a western grid point centered at $-60\degr$ W, $8\degr$ N and red corresponds to a dayside gridpoint centered at $42\degr$ E, $8\degr$ N. The value of $f_{sed}$, which ranges from 0.03 to 6.0, is indicated by the shading of each line, where the lightest shades correspond to the smallest values and darkest shades represent the largest values of $f_{sed}$.}
    \label{fig:OPD-K107}
\end{figure*}

For the \texttt{Virga} cloud modeling in this study, we are only interested in cloud species that are expected to be abundant in these atmospheres and which condense at relatively low pressures (approximately 1 bar).  We identify three cloud species that are likely to be important based on comparing the species' condensation curves to the planets' temperature pressure profiles \citep[Figure~\ref{fig:TCondens}; also see][]{ackerman2001,morley2012}: Mg$_2$SiO$_4$, Al$_2$O$_3$, and TiO$_2$. \color{black}{This assumes that all SiO goes into Mg$_2$SiO$_4$ rather than SiO$_2$ or MgSiO$_3$. The inclusion of SiO$_2$ would require a different modeling framework including kinetic condensation}. \color{black}We plot the optical depths \color{black}{in the Kepler bandpass (shown in Figure 6)} \color{black} for these species as a function of pressure as computed by \texttt{Virga} at a representative location on the planet K2-107b in Figure \ref{fig:OPD-K107}. Although some previous studies of spatially inhomogeneous cloud formation in hot Jupiters also included sulfide clouds \citep[e.g.][]{parmentier2016}, recent microphysical cloud models  \citep[][]{Powell_2019,gao2020} indicate that the high surface energy of sulfide condensates makes homogeneous nucleation unlikely. This conclusion is in good agreement with observational data indicating that the observed cloud opacity for more moderately irradiated hot Jupiters is well-matched by models without sulfide condensates \citep[e.g.][]{chachan2019,gao2020}. We therefore exclude sulfide condensates from our \texttt{Virga} models. Fe is also a potentially important condensate, but like the sulfides previous microphysical modeling suggested that Fe cloud formation proceeds slowly due to Fe's high surface energy \citep[][]{gao2020}. On the other hand, Fe has been considered a major cloud component in many previous works \citep[][]{marley2013,helling2008,marley2015,fortney2005,sudarsky2003}. Therefore, while our nominal \texttt{Virga} models will not include Fe, we will consider another set of simulations where Fe is included to explore its impact on our predicted albedos.  



For the \texttt{CARMA} simulations in this study, we use the same model setup as in \citet{gao2020} where clouds of TiO$_2$, Al$_2$O$_3$, Fe, Mg$_2$SiO$_4$, Cr, MnS, Na$_2$S, ZnS, and KCl are allowed to form. However, as previously discussed, in practice only Mg$_2$SiO$_4$, Al$_2$O$_3$, and TiO$_2$ form in any significant abundance. Of these three species, Mg$_2$SiO$_4$ is particularly unlikely to undergo homogeneous nucleation, as it is not abundant in the vapor phase. Instead, it is the product of a thermochemical reaction between Mg, SiO, and H$_2$O \citep[e.g.][]{Visscher_2010}. Similarly, Al$_2$O$_3$ does not exist in the gaseous phase as other aluminum oxide species will be more stable \citep[e.g.][]{patzer2005}. As in \citet{gao2020}, we allow these two species to heterogeneously nucleate on homogeneously nucleated TiO$_2$ seeds. Though Fe is permitted to both homogeneously and heterogeneously nucleate (on TiO$_2$ seeds), the high surface of Fe prevents significant Fe cloud formation.

\subsection{Computing Optical Albedos with \texttt{PICASO}}\label{sec:picaso}

\begin{figure*}[!tbp]
\centering
\subfloat{\includegraphics[width=0.49\textwidth]{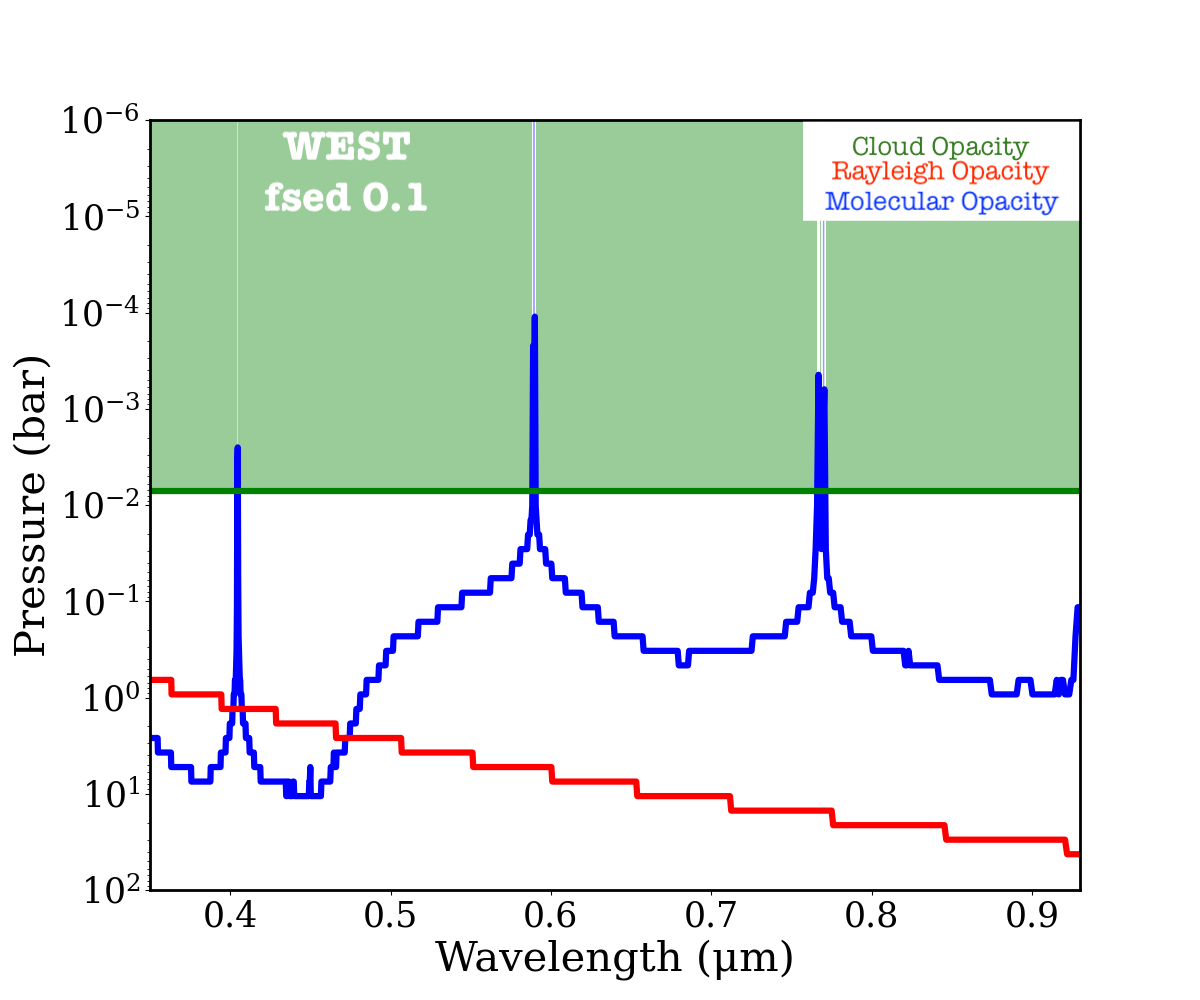}}
\hfill
\subfloat{\includegraphics[width=0.49\textwidth]{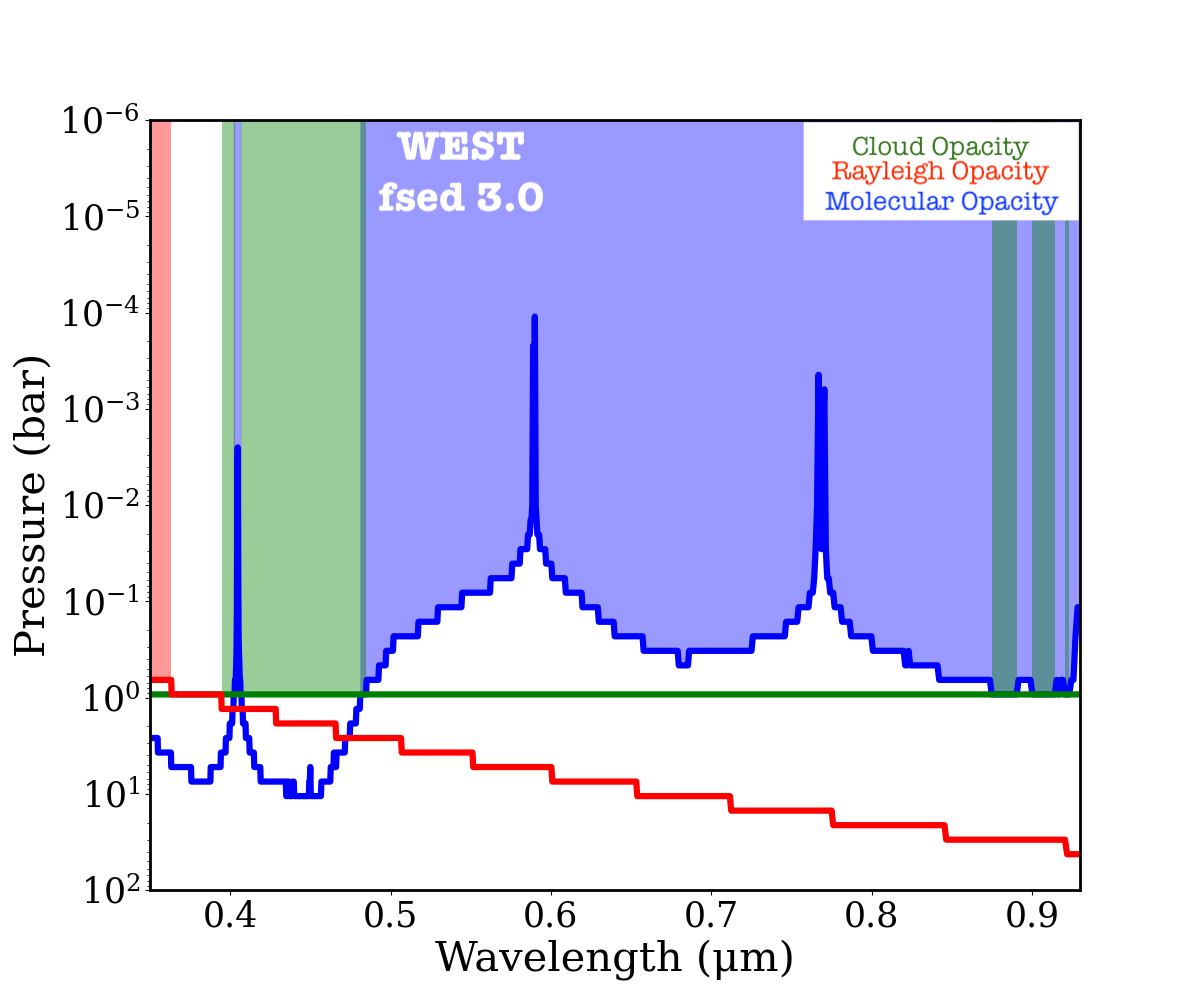}}
\caption{Pressure level (bar) of unit optical depth for Rayleigh scattering (pink), cloud opacity (green), and molecular opacity (black) as a function of wavelength for a single grid point ($-60\degr$ W, $8\degr$ N) in the \texttt{Virga} model of hot Jupiter K2-107b. $f_{sed}$ of 0.1 is shown on the left and 3.0 on the right. The shaded regions indicate the dominant opacity source as a function of wavelength. This is the same grid point shown in Figure \ref{fig:OPD-K107}.
\label{fig:opacity_fsed}}
\end{figure*}

 We use the outputs from the cloud codes \texttt{Virga} and \texttt{CARMA} to calculate the single scattering albedos, asymmetry parameters, and optical depths for each pressure layer at each location in the atmosphere assuming Mie scattering. We then convert these quantities into wavelength-dependent albedos using The Planetary Intensity Code for Atmospheric Scattering Observations \citep[\texttt{PICASO;}][]{picaso}.  This code is  governed by the radiative transfer equation:
 
\begin{equation}
    \begin{gathered}[b]
    I(\tau_i,\mu) = I(\tau_{i+1},\mu) \exp(\frac{\delta \tau_i}{\mu}) - \\ 
    \int_{0}^{\delta \tau_i} S(\tau' \mu) \exp(-\frac{\tau}{\mu}) d\tau'/\mu
    \end{gathered}
    \label{eq:rt1}
\end{equation}
where $I(\tau_i,\mu)$ is the azimuthally averaged intensity emerging from the top of a layer $i$ with opacity $\tau_i$ and outgoing angle $\mu$.  $I(\tau_{i+1})\exp(\frac{\delta \tau_i}{\mu})$ is the incident intensity at the lower boundary of the layer attenuated by the optical depth within the layer $\delta \tau$, and $S(\tau',\mu)$ is the source function integrated over all layers.  The source function has two components: single scattered and multiple scattered radiation integrated over all diffuse angles: 

\begin{equation} 
    \begin{gathered}
    S(\tau',\mu) = \frac{\omega}{4\pi}F_o P_{single} (\mu, -\mu_o) \exp(-\frac{\tau'}{\mu_s}) + \\ 
    \frac{\omega}{2} \int_{-1}^{1}I(\tau',\mu')P_{multi}(\mu,\mu')d\mu',
    \end{gathered}
    \label{eq:rt2}
\end{equation}
where $\omega$ is the single scattering albedo and $F_o$ is the incident flux. $P_{multi}$ and $P_{single}$ describe the phase function of the multiple and single scattering, respectively. $P_{single}$ is an opacity weighted combination of the Rayleigh phase function and a two-term Henyey Greenstein phase function. $P_{multi}$ requires integration over all diffuse angles, for which \texttt{PICASO} uses a $N=2$ Legendre expansion. This approximation alone is inadequate to represent cases with high rates of forward scattering, so \texttt{PICASO} implements the delta-Eddington approximation to scale $g$, $\omega$, and $\tau$ to more accurately capture the forward scattering peak.

\texttt{PICASO} considers the extinction from three opacity sources in order to calculate the geometric albedo as a function of wavelength: molecular absorption, Rayleigh scattering, and scattering by clouds. We show a representative calculation of these three opacity sources as a function of wavelength in Figure \ref{fig:opacity_fsed}.  To describe the phase-dependence, \texttt{PICASO} computes the emergent intensity from the disk at multiple plane-parallel facets, where each has its own incident and outgoing angles. \texttt{PICASO} uses the Chebyshev-Gauss integration method to integrate over all emergent intensities.  We integrate the wavelength-dependent geometric albedo over the \emph{Kepler} response function for each point in our $10\times10$ grid and then integrate again over the dayside hemisphere in order to obtain a geometric albedo that we can compare with the \emph{Kepler} measurements. We have run sensitivity tests that demonstrate a higher resolution grid ($20\times20$ grid) yields comparable results.

\subsection{Effect of Simplifying Model Assumptions}\label{sec:simpl}

In this study we do not consider radiative feedback from clouds, which might affect our albedo predictions.  Clouds can alter the planet's global thermal structure in several ways.  First, they can reduce the amount of heating on the dayside by increasing the planet's Bond albedo.  Second, they can suppress cooling on the nightside by preventing the re-radiation of infrared light to space.  
\cite{parmentier2021} and \cite{roman2021} ran grids of 3D GCM models incorporating radiative feedback from clouds spanning a range of incident fluxes. They found that the presence of reflective clouds on the dayside resulted in lower global temperatures, while the presence of nightside clouds inhibited cooling, causing a $100-200$ K global increase in temperature. With the possible exception of Kepler-7b (see \S\ref{sec:k7}), we expect that such shifts in temperature would not substantially alter the pressures of the cloud decks or reduce their horizontal extent for the planets examined here. 

We also note that in this study, our 3D atmospheric circulation models are decoupled from our cloud models.  While this does enable more flexibility in exploring different cloud species and sedimentation parameters in both \texttt{Virga} and \texttt{CARMA}, cloud formation and transport is ultimately a coupled process between advection, radiation, and chemistry.  Previous studies that couple cloud microphysical models and atmospheric circulation models \citep[e.g.,][]{lee2015, lines2018} suggest that zonal (east/west) and/or meridional (north-south) transport of cloud particles from colder regions of the atmosphere could lead to high cloud particle number densities, even on the hotter dayside, which could in turn enable more cloud nucleation and growth.  These processes will only affect our dayside albedo predictions if the planet in question has large dayside temperature (and hence albedo) gradients with longitude or latitude. We revisit both of these assumptions (radiative feedback and cloud microphysics coupled to circulation) in \S\ref{sec:planets}, where we discuss their implications for individual planets in light of our model results.

\section{Results}\label{sec:results}

The SPARC/MITgcm models indicate that the day-night temperature contrasts for the six planets in our sample vary in magnitude (Figure \ref{fig:tmaps}). As expected, the hottest region on the day side in all six models is located to the east of the substellar point.  This shift is caused by eastward equatorial winds, which transport heat to the planet's nightside \citep[e.g.,][and references therein]{showman2020}.

Kepler-7b has the largest thermal gradient of all the planets in our sample, followed by K2-31b and Kepler-17b; this is because the former has a relatively low surface gravity (approximately 4 m/s$^2$) and the latter two have the shortest orbital periods in the sample

\color{black}{Kepler-7b has the largest thermal gradient of all the planets in our sample, followed by K2-31b and Kepler-17b; this is because the former has a relatively low surface gravity (approximately 4 m/s$^2$) and the latter two have the shortest orbital periods in the sample.} \color{black}Our models indicate that Kepler-7b is also warmer at depth than the other planets in this sample.  This is expected, as Kepler-7b has the lowest surface gravity of the six planets and under hydrostatic equilibrium the gradient of temperature with respect to pressure is related to the inverse of the surface gravity \citep[e.g.,][]{gao2018}.  



\begin{figure}
    \includegraphics[width=3.4in]{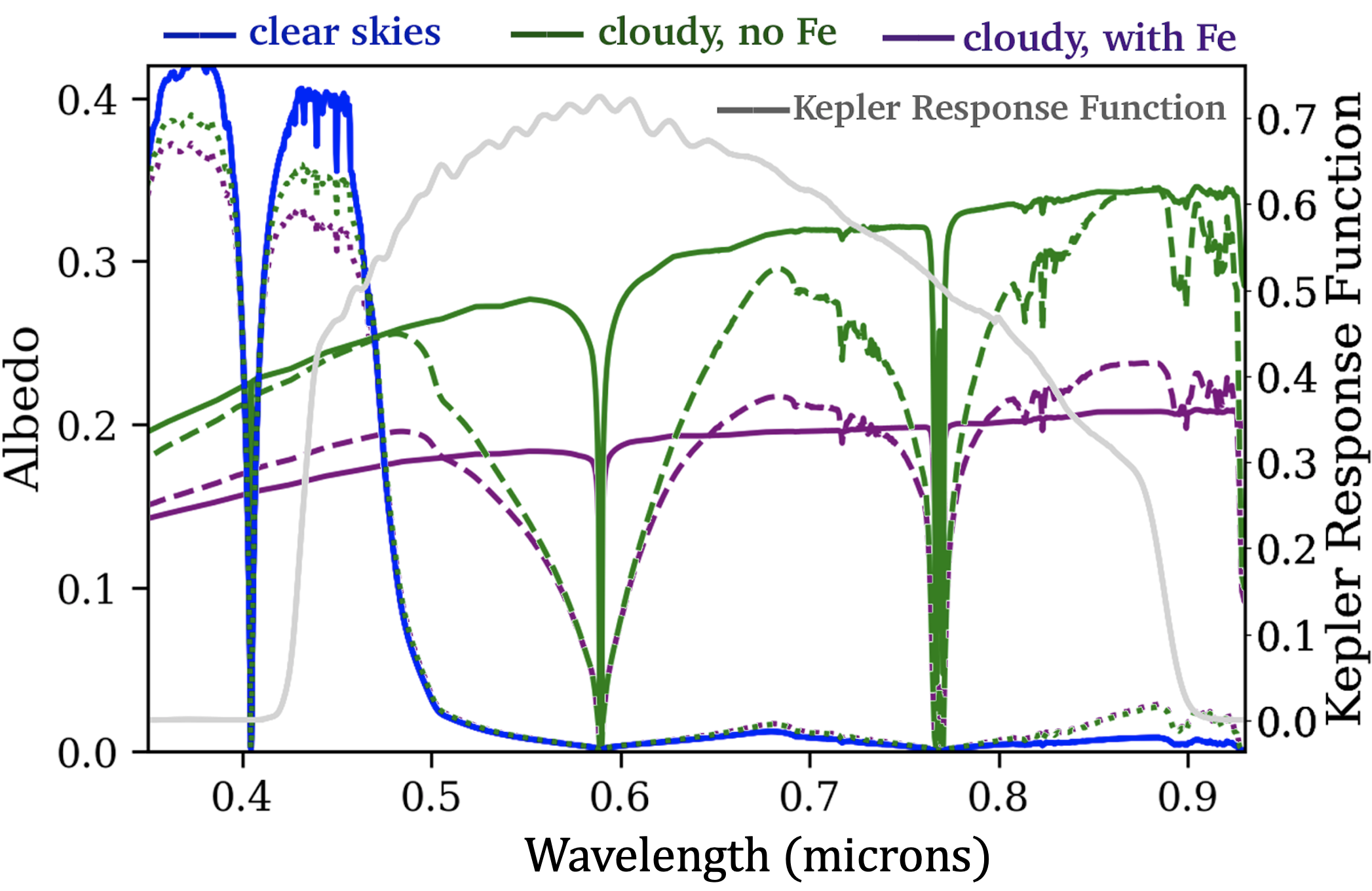}
    \caption{Hemisphere-averaged albedo of K2-107b as a function of wavelength for a clear atmosphere (black) and cloudy atmospheres of varying $f_{sed}$ values (green; 0.03 as a solid line, 0.1 as a dashed line, and 0.3 as a dotted line). Cloudy atmospheres including Fe condensates  are shown in purple. All cloud distributions are computed using \texttt{Virga}. The \emph{Kepler} response function is overplotted in grey.}
    \label{fig:a-K107}
\end{figure}

\begin{figure*}
\centering
\includegraphics[width=0.95\textwidth]{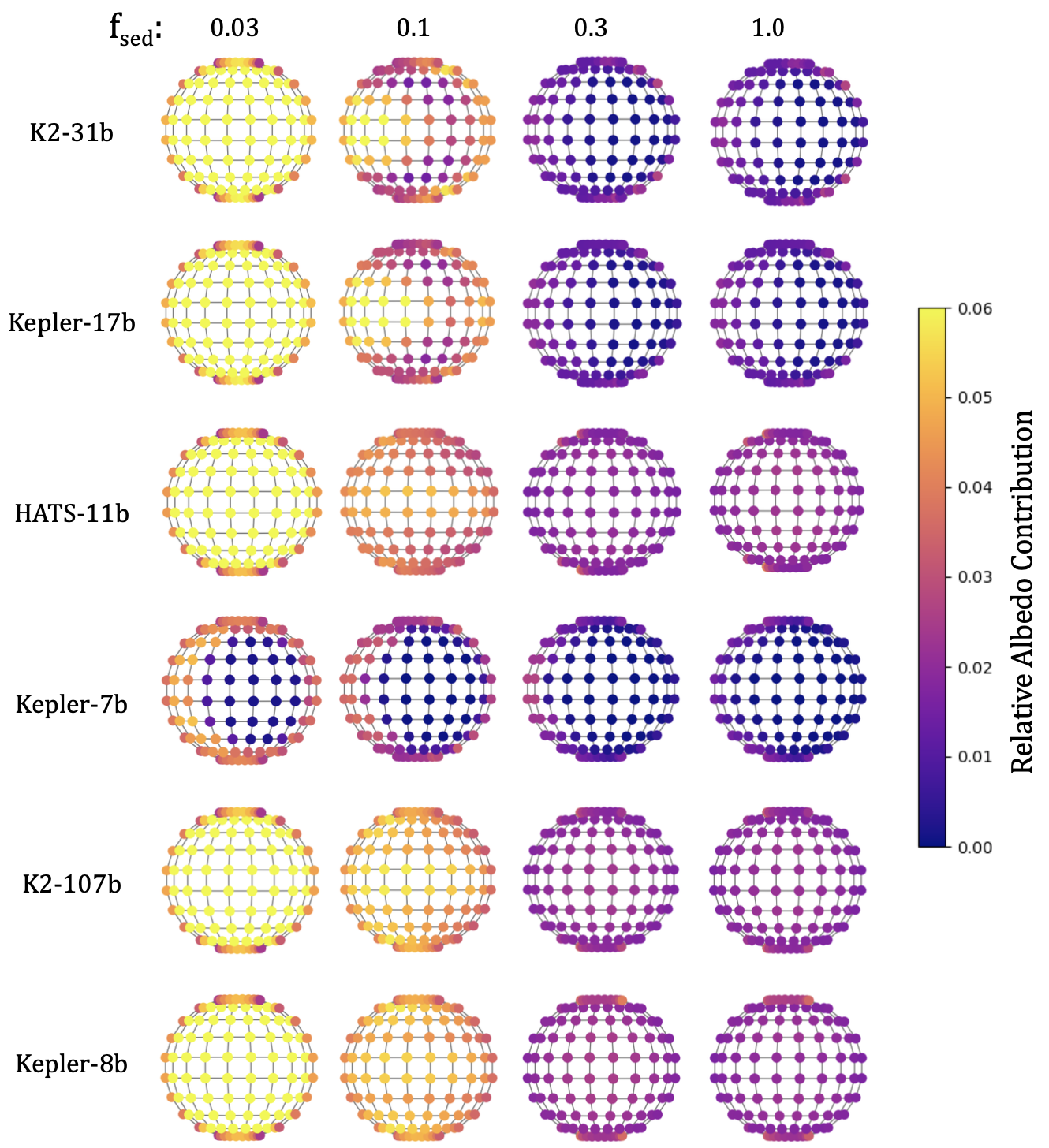}
\caption{Relative dayside albedo contribution at each grid point including appropriate geometric weights.  Planets are sorted by increasing equilibrium temperature from top to bottom, and sorted by increasing $f_{sed}$ (0.03, 0.1, 0.3, and 1.0) from left to right. We omit $f_{sed}$ values of 3.0 and 6.0 from the figure, as the clouds in these models reside below the level of unit molecular albedo, and thus the albedo remains roughly unchanged compared to the $f_{sed}$ = 1.0 case. The relative albedo contributions are all smaller than the face-integrated albedo by roughly a factor of $~\pi$; summing over the points yields the integrated albedo. The models include three cloud species: Mg$_2$SiO$_4$, Al$_2$O$_3$, and TiO$_2$.}
\label{fig:discos}
\end{figure*}

\subsection{\texttt{Virga} Model Results}

In order to determine the effect of the clouds on the albedo, we must first calculate their vertical extent at each location in our model grid.  In our \texttt{Virga} equilibrium cloud models, the vertical extend of the clouds is controlled by our choice of $f_{sed}$.   Taking K2-107b as a representative example, Figure \ref{fig:OPD-K107} indicates that Mg$_2$SiO$_4$ clouds reach a unit optical depth near 1 mbar for the lowest $f_{sed}$ value, 0.03, and near 50 mbar for $f_{sed}=$ 0.1.  These clouds will only contribute to the albedo at wavelengths where they reach optical depth unity at lower pressures than Rayleigh scattering or molecular opacity (Figure \ref{fig:opacity_fsed}).  This means that clouds will have a greater effect at wavelengths where the combined molecular and Rayleigh scattering opacity is lower.  We show the effect of varying $f_{sed}$ on the wavelength-dependent albedo of K2-107b in Figure \ref{fig:a-K107}.  As $f_{sed}$ decreases and the clouds extend to lower pressures, the cloud opacity contributes most to the overall albedo for an increasingly large fraction of the \emph{Kepler} bandpass.

We next examine how the contribution of clouds to the band-integrated albedo varies across the dayside atmosphere.  In Figure \ref{fig:discos}, we show the $10\times10$ grid of albedos in the \emph{Kepler} bandpass for each planet as a function of $f_{sed}$.  We find that the three planets with the greatest temperature variation as a function of longitude (Kepler-7b, K2-31b and Kepler-17b) also have relatively large albedo variations across their dayside atmospheres for low to intermediate $f_{sed}$ values. Once $f_{sed}$ increases above 0.3, the clouds remain confined below the optical depth unity level of molecular absorbers, such that the presence or absence of clouds does not affect the observed albedo. These three planets appear brighter on their western limbs than in the east, in good agreement with the albedo maps derived from the \emph{Kepler} phase curve for Kepler-7b \citep{demory2013} and other planets with comparable equilibrium temperatures \citep{shporer2015}. In contrast to these three planets, HATS-11b, K2-107b, and Kepler-8b all appear to have fairly homogeneous albedos, as expected based on their more homogeneous thermal structures.

\begin{centering}
\begin{deluxetable*}{lccccccccc}[!ht]
\tablecaption{Hemisphere-Averaged Albedos from \texttt{Virga} Models as a Function of $f_{sed}$\tablenotemark{a}}
\tablehead{
 \colhead{Planet} & \colhead{Measured} & \colhead{Clear} & \colhead{0.03} & \colhead{0.1} & \colhead{0.3} & \colhead{1.0} & \colhead{3.0} & \colhead{6.0} & \colhead{0.1, with Fe}}
\startdata
K2-31b & 0.023$\pm$0.002 & 0.015 & 0.404 & 0.123 & \textbf{0.023} & 0.015 & 0.015 & 0.015 & 0.115 \\
Kepler-17b & 0.099$\pm$0.017 & 0.017 & 0.416 & \textbf{0.131} & 0.023 & 0.017 & 0.017 & 0.017 & 0.125 \\
HATS-11b & 0.270$\pm$0.052 & 0.066 & \textbf{0.301} & 0.127 & 0.060 & 0.066 & 0.066 & 0.066 & 0.127 \\
Kepler-7b & 0.194$\pm$0.013 & 0.009 & \textbf{0.064} & 0.034 & 0.015 & 0.010 & 0.009 & 0.009 & 0.030 \\ 
K2-107b & 0.062$\pm$0.010 & 0.065 & 0.333 & 0.154 & \textbf{0.063} & 0.065 & 0.065 & 0.065 & 0.134 \\
Kepler-8b & 0.124$\pm$0.013 & 0.069 & 0.319 & \textbf{0.151} & 0.063 & 0.070 & 0.070 & 0.070 & 0.151
\enddata 
\tablenotetext{a}{Bolded values indicate the simulated \texttt{Virga} albedo that best matches the \emph{Kepler} eclipse observations.}
\end{deluxetable*}\label{table:virga_albedos}
\end{centering}

\begin{figure*}[t]
\centering
\includegraphics[width=0.9\textwidth]{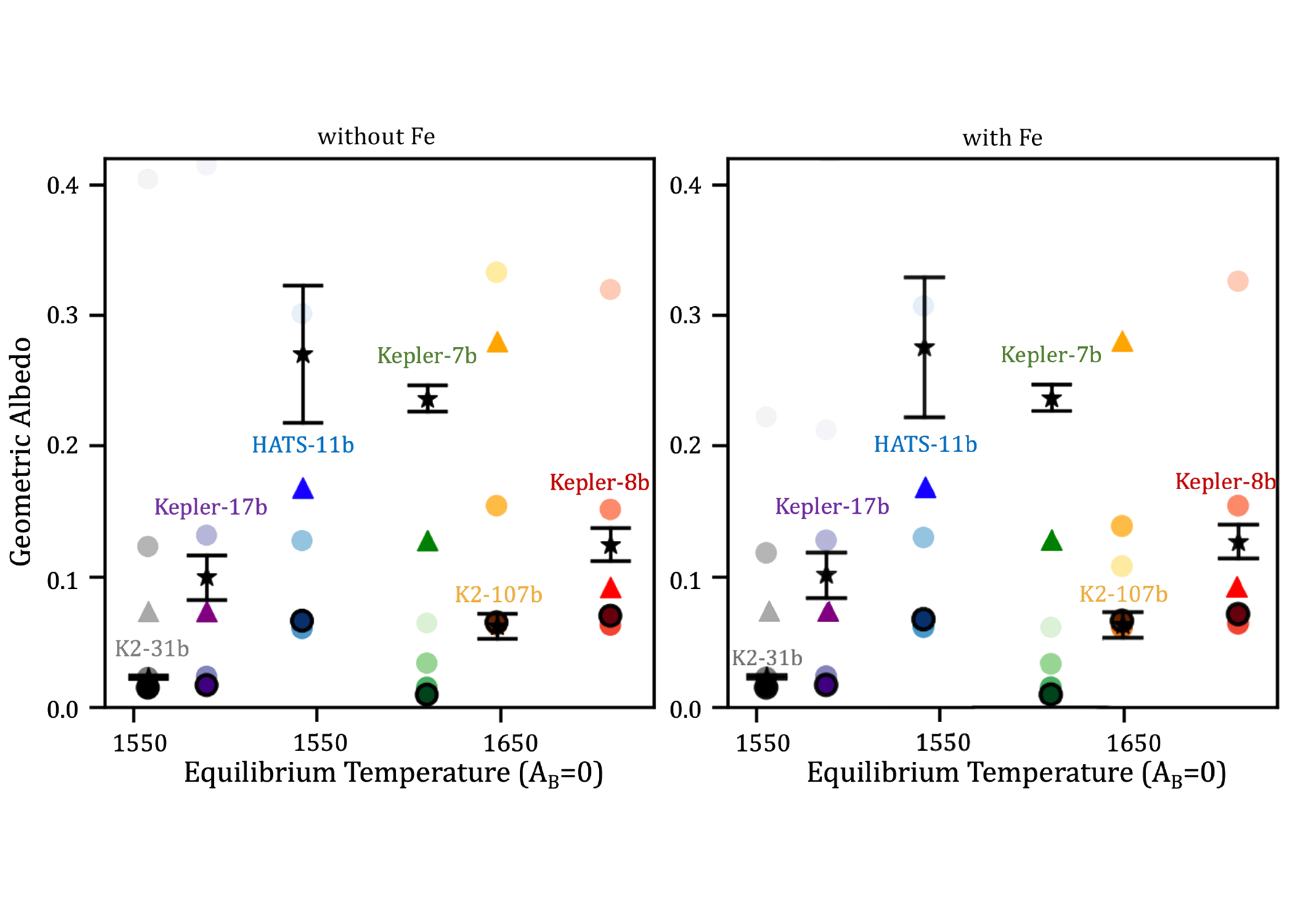}
\caption{Geometric albedo predictions in the \emph{Kepler} optical bandpass compared with published albedo measurements as a function of equilibrium temperature. Albedo predictions from the full-resolution \texttt{Virga} models are shown as circles where the shading indicates the $f_{sed}$ value, going from 0.03 (light) to 6.0 (dark). \emph{Kepler} albedo measurements are shown as stars while the predicted albedos from the two-zone CARMA models are shown as triangles. Left panel excludes Fe in \texttt{Virga} calculations, while the right panel includes Fe condensates.}
\label{fig:summary}
\end{figure*}

Lastly, we compare the hemisphere-averaged dayside albedo in the \emph{Kepler} bandpass as a function of $f_{sed}$ to the measured dayside albedo for each planet (Figure \ref{fig:summary}). We find that K2-31b and K2-107b are best described by models with large values of $f_{sed}$ or (equivalently) clear atmospheres, indicating that any reflective clouds present in these atmospheres do not extend above the level of unit molecular opacity. Kepler-17b and Kepler-8b are best-matched by models with intermediate $f_{sed}$ values, implying that their albedos are moderately enhanced by cloud opacity. HATS-11b is best matched by models with low $f_{sed}$ values, corresponding to a high, vertically extended reflective cloud layer spanning a wide range of longitudes. No $f_{sed}$ value is able to reproduce the high observed albedo for Kepler-7b, which our models indicate is too warm for bright Mg$_2$SiO$_4$ clouds to form over much of the day side. Instead, our models indicate that the dayside of Kepler-7b is dominated by deep Al$_2$O$_3$ clouds at pressures of around a bar. The addition of iron clouds do not greatly change our results except for decreasing the albedos of low $f_{sed}$ models, as iron clouds sink below the photosphere for higher $f_{sed}$'s.  We list the observed albedos and the predicted albedos as a function of $f_{sed}$ in Table \ref{table:virga_albedos}.  We find that the hemisphere-integrated albedo over the \emph{Kepler} bandpass is very sensitive to the assumed value of $f_{sed}$.

\subsection{Comparison to \texttt{CARMA} Microphysical Models}\label{sec:carma_comparison}

We find that dayside albedos can vary significantly depending on the assumed sedimentation efficiency (e.g., Figure \ref{fig:discos}). We therefore use these maps to divide each planet into two zones, and then utilize the more computationally demanding \texttt{CARMA} microphysical cloud model to solve for the vertical extent of the clouds and corresponding albedo in each zone. 

Our \texttt{Virga} albedo maps indicate that a subset of the planets in our sample are likely to have spatially inhomogenous Mg$_2$SiO$_4$ clouds located near their western limbs. Although it would be computationally prohibitive to run a separate \texttt{CARMA} model for each point in the $10\times10$ grid, we can nonetheless capture this cloud structure using a more computationally tractable two-zone model. We define a threshold value in longitude for each planet based on the albedo distributions found by \texttt{Virga} as shown in Figure \ref{fig:discos} and in Table \ref{table:carma_albedos}. We determine the longitude range defining the two zones (a western zone and a dayside zone) as the division that yields the greatest difference in albedo at mid-latitudes as determined from the \texttt{Virga} albedo maps. Occasionally the western zone includes a limb region in the east; see Table \ref{table:carma_albedos} for a list of the zone definitions for each planet. The resulting zonally-averaged pressure-temperature profiles are plotted in Figure \ref{fig:TCondens}. We run a separate \texttt{CARMA} model for each zone, and also run \texttt{Virga} models for the same zones in order to facilitate comparisons between the two models.  In the discussion below, we limit our comparisons to these two-zone \texttt{Virga} models unless otherwise noted.

\begin{figure*}
    \centering
    \includegraphics[width=1.0\textwidth]{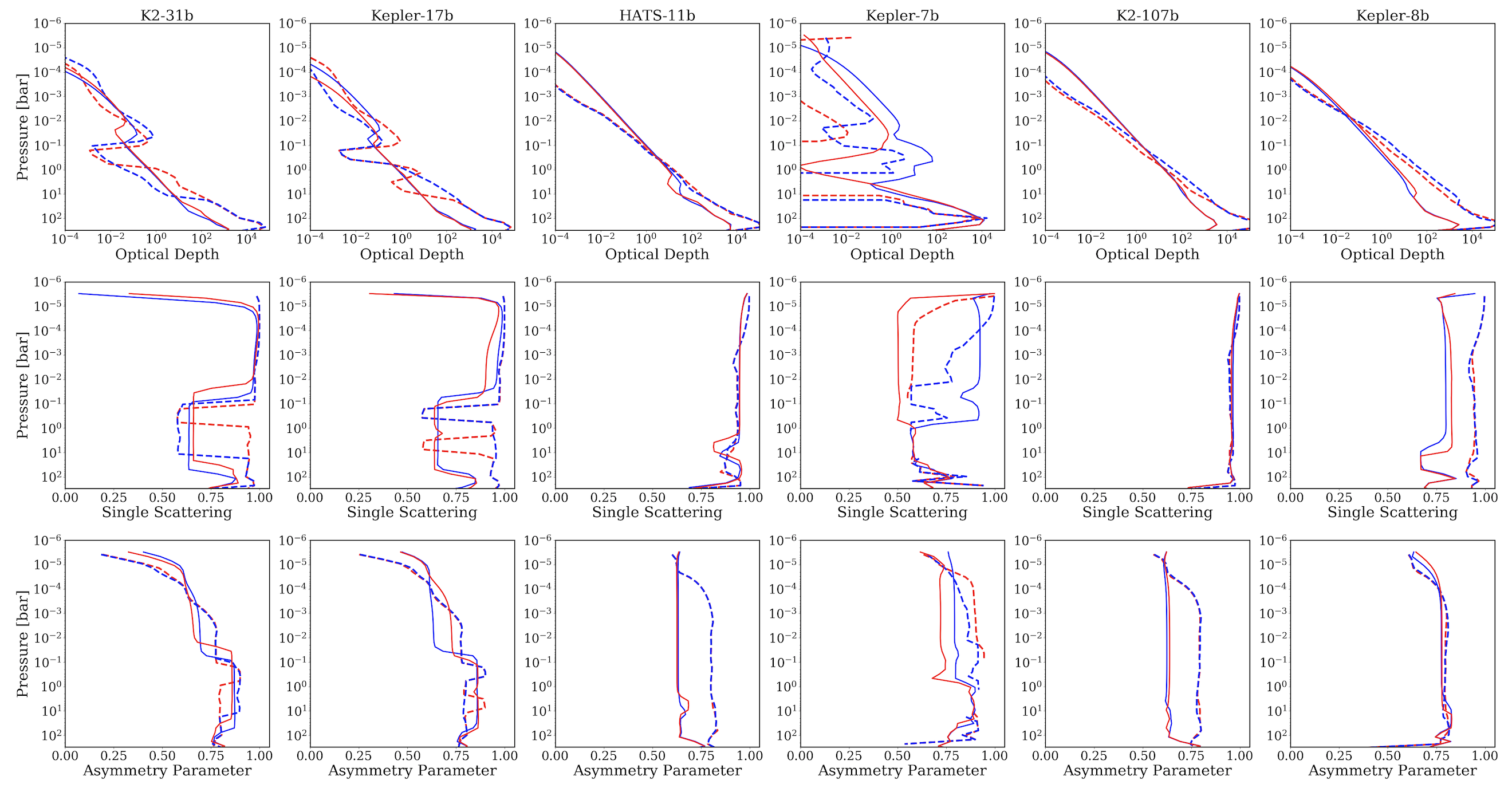}
    \caption{Optical depth (left), single scattering albedo (center), and asymmetry parameter (right) as a function of pressure for the two-zone \texttt{Virga} (dashed; $f_{sed}$ fixed to 0.1) and \texttt{CARMA} (solid) models, calculated by dividing the dayside hemisphere into a dayside (red) and western (black) zone (see \S\ref{sec:carma_comparison}). Each row corresponds to a different planet, sorted left-to-right by increasing equilibrium temperature.}
    \label{fig:opd_w_g}
\end{figure*}

\begin{figure*}
    \centering
    \includegraphics[width=0.85\textwidth]{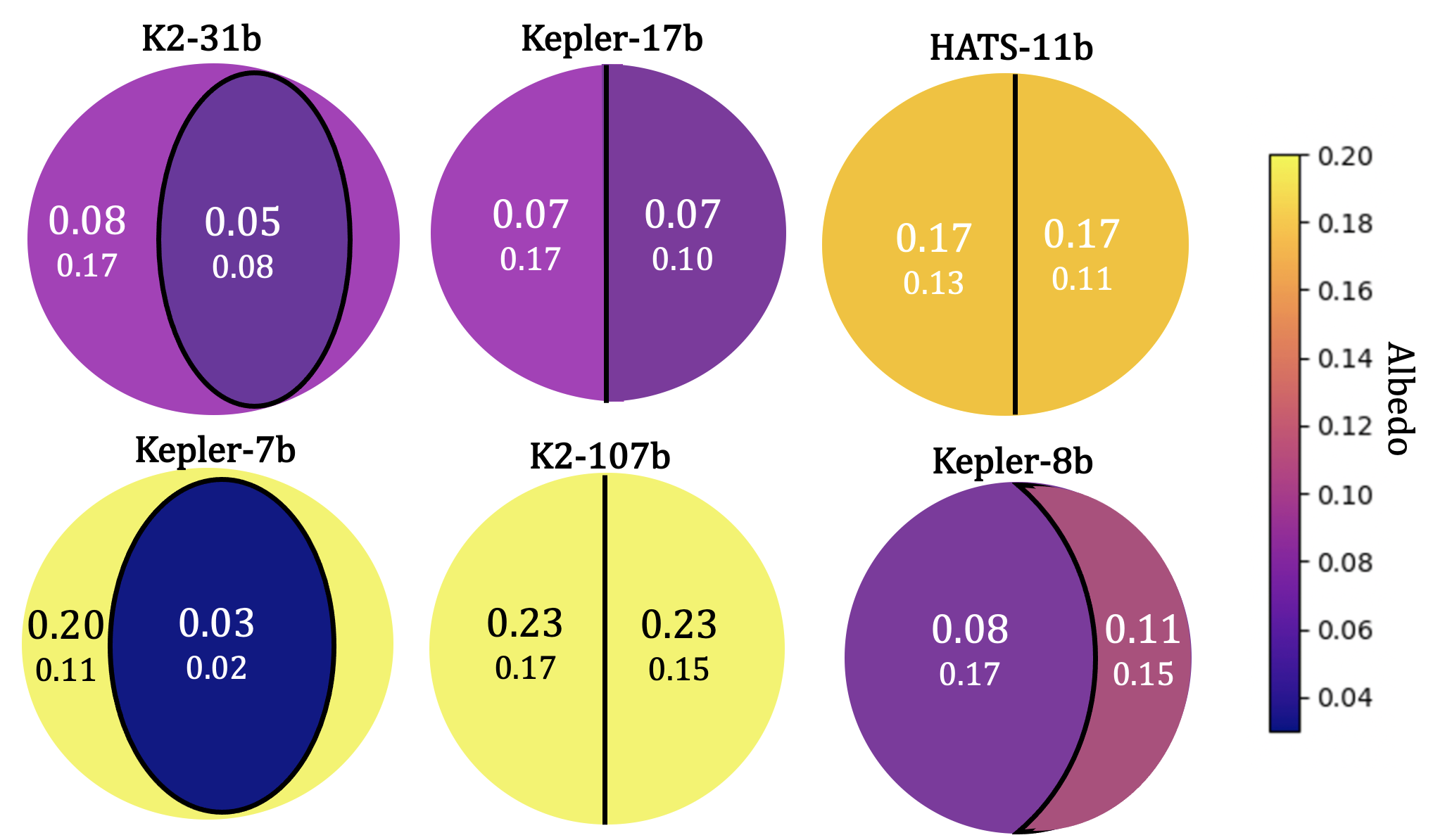}
    \caption{Dayside and western zone albedos from the \texttt{CARMA} models integrated over the \emph{Kepler} bandpass.  \texttt{Virga} model albedos with $f_{sed}$ equal to 0.1 are listed below for comparison.  Both sets of models are calculated in two zonally averaged regions, defined in Table \ref{table:carma_albedos}. Planets are sorted from left to right and top to bottom by increasing equilibrium temperature.}
    \label{fig:disco_CARMA}
\end{figure*}

When comparing the two models, it is important to note that they predict distinct particle compositions.  While \texttt{Virga} assumes that particles nucleate homogeneously, \texttt{CARMA} predicts that the Mg$_2$SiO$_4$ condensates will nucleate heterogeneously onto TiO$_2$ cores (see \S\ref{sec:cloudcomp}). We find that accounting for this TiO$_2$ core when we calculate the single scattering albedos for the Mg$_2$SiO$_4$ particles in our \texttt{CARMA} models results in a lower single scattering albedo than when we repeat the calculation for pure Mg$_2$SiO$_4$ particles, which return single scattering albedo profiles of greater than 0.96. For comparison, we refer the reader to Figure \ref{fig:opd_w_g}; a single value for single scattering albedo is not representative of heterogeneous particles, which are sensitive to the core mass fraction of TiO$_2$. However, even after accounting for this effect we find that the species-averaged single scattering albedo shown in Figure \ref{fig:opd_w_g} for our \texttt{CARMA} models with heterogeneous particles is still comparable to the \texttt{Virga} models with $f_{sed}$ equal to 0.1 (Figure \ref{fig:disco_CARMA}).

When we calculate the hemisphere-integrated albedos for these planets from the \texttt{CARMA} models we find that they generally lie within the range of \texttt{Virga} predictions for varying $f_{sed}$ (see Figure \ref{fig:summary} and Table \ref{table:carma_albedos}).  \textcolor{black}{In Figure \ref{fig:summary},} we show the hemisphere-integrated albedos from the full-resolution \texttt{Virga} model from \S\ref{sec:virga} rather than the two-zone model discussed in this section. 

We find that K2-107b and HATS-11b appear to have the brightest hemisphere-integrated \texttt{CARMA} albedos, with values consistent with those of the lowest $f_{sed}$ ($0.03-0.1$) \texttt{Virga} models.  Kepler-8b, Kepler-17b, and K2-31b have somewhat lower predicted \texttt{CARMA} albedos, more comparable to the $f_{sed}$ of $0.3$ \texttt{Virga} models. Kepler-7b is a notable exception, as our \texttt{CARMA} model predicts an albedo that is higher than that of the smallest $f_{sed}$ \texttt{Virga} model.

\begin{centering}
\begin{deluxetable*}{lcccccc}[!t]
\tablecaption{Hemisphere-Averaged Dayside Albedos from \texttt{CARMA} Models}
\tablewidth{0pt}
\tablehead{
 \colhead{Planet} & \colhead{Measured} & \colhead{Clear} & \colhead{West} & \colhead{Day} & \colhead{Day Zone\tablenotemark{1}} & \colhead{Integrated}}
\startdata
K2-31b & 0.023$\pm$0.002 & 0.015 & 0.085 & 0.054 & $[-17\degr W, 51\degr E]$ & 0.073 \\
Kepler-17b & 0.099$\pm$0.017 & 0.017 & 0.074 & 0.069 & $[0\degr E, 90\degr E]$ & 0.073 \\
HATS-11B & 0.270$\pm$0.052 & 0.066 & 0.169 & 0.165 & $[51\degr E, 90\degr E]$ & 0.168 \\
Kepler-7b & 0.194$\pm$0.013 & 0.009 & 0.199 & 0.034 & $[-38\degr W, 51\degr E]$ & 0.128 \\ 
K2-107b & 0.062$\pm$0.010 & 0.065 & 0.233 & 0.229 & $[51\degr E, 90\degr E]$ & 0.280 \\
Kepler-8b & 0.124$\pm$0.013 & 0.069 & 0.082 & 0.109 &  $[51\degr E, 90\degr E]$ & 0.092
\enddata 
\tablenotetext{1}{Eastern/day zone boundaries are listed in the table. The western/limb zone is defined as the region excluded by the dayside definition.}
\end{deluxetable*}\label{table:carma_albedos}
\end{centering}


\section{Discussion}\label{sec:discussion}

\subsection{Comparison to Previous Modeling Efforts}

We can compare our results to generic grids of cloudy GCM models in the published literature. \color{black}{First, we note that our GCM outputs for Kepler-7b roughly agree with those shown in \citet{oreshenko2016}, and the general distribution of our silicate and corundum clouds agrees with the modeling results of \citet{roman2019}}. 

\color{black}\citet{roman2021} investigated planetary albedos using a grid of GCMs with varying irradiation temperature and surface gravity. The closest equivalent models in their grid are for a planet with an irradiation temperature of either 2500 K (equilibrium temperature of 1500 K) or 2750 K (equilibrium temperature of 1700 K) and a surface gravity of 10 m/s$^2$.  We compare to their nucleation-limited models, which also exclude iron and sulfide condensates.   

The models presented in \citet{roman2021} assume a fixed pressure-dependent particle size for the clouds, with a size of 0.1 $\mu$m at the top of the atmosphere that increases exponentially with increasing pressure for pressures greater than 10 mbar.  Since their equilibrium  cloud models do not solve for the vertical extent of the cloud layers, they present two cases corresponding to compact (cloud tops limited to 1.4 scale heights above the cloud base) and vertically extended (cloud top pressure of 0.1 mbar) cloud layers. Unlike our models, they account for radiative feedback from these clouds when solving for the temperature structure of the atmosphere. These models indicate that compact cloud layers will result in relatively low and uniform dayside albedos, in good agreement with our results (Figure \ref{fig:discos}).  Their vertically extended cloud models exhibit a range of optical spherical albedos between $0.2-0.3$, with lower values for the higher temperature model as the reflective silicate clouds become increasingly confined to the cooler western region of the atmosphere.  This also agrees with the qualitative picture from our models, which span an equivalent range of albedos.  They  conclude that their optical albedos for the nucleation-limited case are dominated by silicate cloud particles, consistent with our conclusions here.  

\citet{parmentier2016} and \citet{parmentier2021} considered GCMs spanning a wide range of equilibrium temperatures.  In \cite{parmentier2016}, the clouds are post-processed (i.e., they do not include cloud radiative feedback in the GCM models), and they model the condensation of a wide range of cloud species.  These models predict that MgSiO$_3$ and CaTiO$_3$ clouds should dominate the dayside optical albedo for planets with equilibrium temperatures between $1500-1700$ K.  In \cite{parmentier2021} they incorporate cloud radiative feedback for the condensation of a single cloud species, MnS, but the treatment of silicate clouds is similar to \citet{parmentier2016}.  For that reason, we focus on \cite{parmentier2016} for our comparison.  In this study they assume a small fixed particle size distribution centered at 0.1 $\mu$m and a cloud top pressure of 1 microbar, which yields geometric albedos greater than 0.5 in the \emph{Kepler} bandpass for planets of approximately 1500 K equilibrium temperature. This value is much higher than both our albedos and those reported by \cite{roman2021}, and is most likely due to the very low cloud top pressure assumed in these models. They invoke a cold trap for silicates to reduce the albedo, while we predict that changing the cloud vertical extent can achieve a similar difference in observable albedo.

\subsection{Comparison to Published Kepler Albedos}\label{sec:planets}

Our results show that, with the exception of Kepler-7b, it is possible to match the observed optical geometric albedos for all of the planets in our sample using either \texttt{Virga} or \texttt{CARMA} models. However, no single model (\texttt{Virga} at a fixed $f_{sed}$ or \texttt{CARMA}) can explain the observed albedos of all six planets. Although the albedo predictions from the \texttt{CARMA} models are broadly consistent with those of the \texttt{Virga} models, neither model is able to reproduce or explain the observed planet-to-planet variations in dayside albedo. This suggests that accounting for differences in equilibrium temperature, host star spectral type, surface gravity, and rotation rate alone are not sufficient to capture the observed diversity of hot Jupiter albedos in this temperature range. We discuss our results in more detail on a planet-by-planet basis below. 

\subsubsection{Kepler-7b: Spatially Resolved Clouds}\label{sec:k7}

Our \texttt{CARMA} and \texttt{Virga} models both indicate that Kepler-7b's relatively bright dayside albedo is dominated by reflection from the region near the western limb, in good agreement with spatially resolved albedo constraints from phase curve observations \citep{demory2013,heng2020}.  This underscores the importance of using spatially resolved cloud models for tidally locked hot Jupiters.  We note that there is some tension between our model predictions and the observational data, as fits to Kepler-7b's optical phase curve indicate that the bright reflective western zone extends as far as $10\pm6\degr$ west of the substellar point \textcolor{black}{\citep[][]{munoz2015, heng2020}}.  Our \texttt{Virga} models predict that the atmosphere will only be cool enough for Mg$_2$SiO$_4$ clouds to condense in the two westernmost longitude bins (extending from the terminator to approximately 38$\degr$ west of the substellar point; see Figure \ref{fig:discos} and Table \ref{table:carma_albedos}).  This is likely why our models under-predict Kepler-7b's optical geometric albedo.  

We consider two possible explanations for this discrepancy.  Zonal  transport of cloud particles from the western limb region could increase the albedo in adjacent longitudes where the atmosphere is otherwise too warm for them to condense (see \S\ref{sec:simpl}).  However, our models for Kepler-7b prefer small cloud particles with a large vertical extent; these small particles might have a relatively short lifetime in the hotter substellar region of the dayside atmosphere.  Instead, perhaps small particles transported meridionally could nucleate and grow in bands at high latitudes \citep{lines2018}.  Ultimately, this planet would be an interesting test case for microphysical transport models \citep[e.g.,][]{lee2015,lines2018}, which can explicitly quantify the timescales of these processes and predict the resulting horizontal distribution of cloud particles. 

Alternatively, if Kepler-7b's dayside is cooler than predicted by our GCM, Mg$_2$SiO$_4$ would be able to condense over a wider range of longitudes. Our GCMs do not account for reflectivity from clouds when calculating the effect of incident starlight on the dayside atmosphere; this effect might reduce the magnitude of dayside heating and result in globally lower temperatures \citep{lines2018,roman2019,roman2021}. However, if the clouds extend over a significant fraction of the planet's nightside it could result in net global warming, as they would act to reduce the amount of energy that can be radiated to space in this region \citep{roman2019,roman2021,parmentier2021}.  For Kepler-7b, whose clouds extend over much of the western hemisphere, it is unclear which of these two competing effects would dominate. 
\textcolor{black}{These explanations assume that the dayside coverage area is the most significant limiting factor on the brightness of the dayside-integrated albedo. However, it is also important to consider factors that might increase the brightness of the cloudy region, including a larger vertical extent for the clouds or brighter cloud particles. We discuss how porous particles may increase HATS-11b’s dayside albedo in Section 4.2.3; this same explanation might also apply to Kepler-7b.}. 

\subsubsection{K2-31b and K2-107b Do Not Host High Altitude Reflective Cloud Layers}

The observed albedos of K2-31b and K2-107b are relatively low, and are well-matched by \texttt{Virga} models with clear skies and/or deep clouds (i.e., those with opacities dominated by molecular absorption). For K2-31b, the contribution of clouds to the albedo is negligible for $f_{sed}$ of 0.3 and larger.  For K2-107b, we obtain a comparable result for $f_{sed}$ of 1.0 and larger.  This suggests that both of these planets may have relatively efficient sedimentation (deep clouds), or alternatively that they have relatively little condensable material in their atmospheres (perhaps corresponding to a relatively low atmospheric metallicity). If the lack of clouds is due to efficient sedimentation, this would appear to contradict predictions from our microphysical \texttt{CARMA} models, which track the sedimentation of cloud particles explicitly and predict albedos that are a factor of two or more higher than the observed values for these two planets.  

Our \texttt{CARMA} models utilize vertical mixing rates calculated from our GCMs.  If these mixing rates are overestimates of the true values, we might expect any clouds near the day-night terminator on these planets to also be relatively compact.  If this is the case, the transmission spectra of these two planets should show relatively strong absorption features.  Although K2-31b has a high surface gravity and is therefore a more challenging target for transmission spectroscopy, K2-107b might be accessible to future space telescopes like \emph{JWST}.  More broadly, the sedimentation rates calculated from GCMs and \texttt{CARMA} could be tested with comparisons to transmission spectra from ongoing surveys \citep[][]{sing2016,crossfield2017,fu2017}. If \texttt{CARMA} models underestimate the sedimentation efficiency for other planets, the disagreement should be detectable in these data, which are very sensitive to the vertical distribution of cloud particles near the limb. To date, most planets appear to be well-matched by \texttt{CARMA} model predictions \citep[e.g.][]{chachan2020,gao2020}, indicating that such model-data disagreements may be relatively rare. 

\subsubsection{HATS-11b, K2-31b, and K2-107b Have Reflective Dayside Clouds}\label{hats}

The measured \emph{Kepler} albedo for HATS-11b ($0.27\pm0.05$) is brighter than that of Kepler-7b, in good agreement with our model predictions.  This planet is cooler than Kepler-7b, with a smaller day-night temperature gradient.  As a result, our models predict a global reflective dayside cloud layer for small $f_{sed}$ values, and our hemisphere-integrated \texttt{Virga} albedo for the $f_{sed}$ of 0.03 model is within $1\sigma$ of the measured value. This is an unusually small value of f$_{sed}$, compared to the other planets' best-fit value, while more typical values of 0.1 and larger underestimate the observation. The \texttt{CARMA} model albedo is somewhat lower, but is still within $2\sigma$ of the observed value. If we wish to adapt our models to better match this planet's high dayside albedo, it likely would require increasing the predicted cloud opacity, such as by increasing the porosity of the cloud particles \citep{samra2020}.  Although increasing the atmospheric metallicity might also increase the cloud opacity, published models for other planets indicate that there is not a simple scaling between these two quantities \citep[e.g.][]{morley2013,gao2018}, and increasing the metallicity will also affect the global thermal structure \citep[e.g.][]{kataria2015}. 

Our model predictions for Kepler-17b and Kepler-8b are also in reasonable agreement with the observed albedos. In both cases, an $f_{sed}=0.1$ \texttt{Virga} model slightly overestimates the albedo while our \texttt{CARMA} model slightly underestimates the albedo. This may indicate that moderately bright/cloudy worlds have moderate sedimentation efficiencies.  For Kepler-8b, both \texttt{CARMA} and \texttt{Virga} models predict that the planet will have relatively uniform cloud coverage in both latitude and longitude; we therefore do not need to consider further spatial variations in cloud number density and particle size.  For Kepler-17b, the $f_{sed}=0.1$ \texttt{Virga} model predicts an albedo gradient across the dayside atmosphere, but this gradient appears to be localized near the equatorial (low latitude) region of the atmosphere.  Our \texttt{CARMA} model predicts a relatively uniform albedo across the two zones, but this may be biased by our inability to resolve latitudinal gradients in the simplified two-zone model.

\section{Conclusions}\label{sec:conclusions}

Optical secondary eclipse measurements made by \emph{Kepler} reveal a wide range of geometric albedos for hot Jupiters with equilibrium temperatures between 1550-1700 K. We combine 3D general circulation models with both equilibrium (\texttt{Virga}) and microphysical (\texttt{CARMA}) cloud models to explore whether 3D effects can explain these observations. We find that the predicted albedos from our \texttt{Virga} models are very sensitive to the assumed sedimentation efficiency ($f_{sed}$).  We can compare these albedo predictions to results from our \texttt{CARMA} model, which use mixing rates calculated from the GCM models to predict the vertical extent and particle size distributions of the clouds. We find that while the hemisphere-integrated \texttt{CARMA} albedos generally agree with the range of albedos predicted by \texttt{Virga}, there is no single $f_{sed}$ value that consistently matches the \texttt{CARMA} predictions. 

When we compare these model predictions to the measured \emph{Kepler} albedos for each of the six planets in our sample, we find that the albedos of K2-31b and K2-107b are best matched by models that are either cloud-free or have very deep compact cloud layers (large $f_{sed}$ values). Kepler-8b and Kepler-17b's optical albedos can be matched by moderately cloudy models  ($f_{sed}$ greater than 0.3). Both \texttt{Virga} and \texttt{CARMA} tend to under-predict the dayside albedos of the two most reflective planets in our sample, HATS-11b and Kepler-7b, which are best matched by \texttt{Virga} models with reflective Mg$_2$SiO$_4$ clouds extending to very low pressures ($f_{sed}$ = 0.03); our \texttt{CARMA} model for HATS-11b predicts a slightly lower albedo value than the brightest \texttt{Virga} model, while our \texttt{CARMA} model for Kepler-7b predicts a slightly higher albedo value than \texttt{Virga}.  Although HATS-11b has relatively uniform cloud coverage across the dayside, it is possible that other factors (such as a low particle porosity) might increase the dayside cloud opacity beyond the values predicted by our models.  

Our models predict that the observed albedo of Kepler-7b should be lower than that of HATS-11b, in good agreement with the observations. Although a bright reflective cloud layer forms in the westernmost region of the dayside atmosphere, most dayside longitudes in Kepler-7b's atmosphere are too warm for Mg$_2$SiO$_4$ to condense, resulting in a lower hemisphere-averaged dayside albedo. Empirical constraints on the horizontal extent of the western cloudy region from phase curve observations indicate that it extends farther east than predicted by our models, hinting that a more detailed study of the planet that couples cloud microphysics and dynamics is required.

We conclude that the sample of optical albedos measured by \emph{Kepler} represents a rich source of information for 3D cloud models, and that there is no single explanation for the observed diversity of albedos for the planets considered in this study.  Future studies leveraging the large sample of transmission spectra of hot Jupiters could provide complementary constraints on the typical sedimentation efficiencies their atmospheres, while additional complementary modeling studies exploring the coupled effects of atmospheric dynamics and cloud microphysics, as well as an exploration of the micro-porosity of cloud particles, would help to further illuminate the relative importance of these processes in explaining the high albedos of the brightest planets in our sample.

\acknowledgments
\color{black}{We thank the anonymous reviewer for an interesting report that greatly helped improve the paper.} \color{black}P. Gao acknowledges support from NASA through the NASA Hubble Fellowship grant HST-HF2-51456.001-A awarded by the Space Telescope Science Institute, which is operated by the Association of Universities for Research in Astronomy, Inc., for NASA, under contract NAS5-26555.

\software{numba \citep{numba}, pandas \citep{mckinney2010data}, bokeh \citep{bokeh}, NumPy \citep{walt2011numpy}, IPython \citep{perez2007ipython}, Jupyter, \citep{kluyver2016jupyter}, Virga \citep{virga}, PICASO \citep{picaso}}

\vspace{15mm}

\nocite{*}
\bibliography{references}{}
\bibliographystyle{aasjournal}

\end{document}